\documentclass[a4paper]{article}

\usepackage{geometry}
\usepackage{subfigure}
\usepackage{caption}
\usepackage{verbatim}
\usepackage[table,xcdraw]{xcolor}
\usepackage{enumitem, kantlipsum}
\usepackage{fancyvrb}
\usepackage{longtable}
\usepackage{lscape}
\usepackage{amssymb,amsmath}
\usepackage{graphicx}
\usepackage{float}
\usepackage{multirow}
\usepackage{tabularx}
\usepackage{marginnote}

\usepackage{listings}
\usepackage{xcolor}

\definecolor{mygray}{gray}{0.6}

\lstdefinelanguage{eQASM}
{
  sensitive=false, 
  keywords={},
  otherkeywords={ 
  |
  },
  morekeywords=[2]{always, never, eq, ne, eqz, nez, lt, ltz, le, gt, ge, gez, ltu, leu, gtu, geu, carry, notcarry},
  morekeywords=[3]{qwait, qwaitr, SMIS, SMIT},
  morekeywords=[4]{bs},
  morekeywords=[5]{MeasZ, X, Y, CNOT, X180, Xm180, X90, Xm90, Y180, Ym180, Ym90, Y90, CZ, T, H, S, Z, QNOP},
  morekeywords=[6]{nop, stop, ldi, ldui, add, sub, addc, subc, and, or, xor, not, nand, nor, xnor, fmr, br, cmp, fbr, goto, sd},
  morecomment=[l]{\#}, 
}

\definecolor{eclipseBlue}{RGB}{42,0.0,255}
\definecolor{eclipseGreen}{RGB}{63,127,95}
\definecolor{eclipsePurple}{RGB}{127,0,85}
\definecolor{sublimegreen}{RGB}{112,200,0}
\lstnewenvironment{eQASM}%
{\lstset{language=eQASM}}
{}
\let\OLDthebibliography\thebibliography
\renewcommand\thebibliography[1]{
  \OLDthebibliography{#1}
  \setlength{\parskip}{0pt}
}

\title{Quantum Computer Architecture:  \\ Towards Full-Stack Quantum Accelerators}

\author{K.~Bertels, A.~Sarkar,  A.A.~Mouedenne,\\ T.~Hubregtsen, A.~Yadav, A.~Krol, I.~Ashraf\\ Quantum Computer Architecture lab\\Delft University of Technology, Netherlands }

\begin{document}

\maketitle

    

\begin{abstract}
This paper presents the definition and implementation of a quantum computer architecture to enable creating a new computational device - a quantum computer as an accelerator.
A key question addressed is what such a quantum computer is and how it relates to the classical processor that controls the entire execution process.
In this paper, we present explicitly the idea of a quantum accelerator which contains the full stack of the layers of an accelerator.
Such a stack starts at the highest level describing the target application of the accelerator.
The next layer abstracts the quantum logic outlining the algorithm that is to be executed on the quantum accelerator.
In our case, the logic is expressed in the universal quantum-classical hybrid computation language developed in the group, called OpenQL, which visualised the quantum processor as a computational accelerator.
The OpenQL compiler translates the program to a common assembly language, called cQASM, which can be executed on a quantum simulator.
The cQASM represents the instruction set that can be executed by the micro-architecture implemented in the quantum accelerator.  
In a subsequent step, the compiler can convert the cQASM to generate the eQASM, which is executable on a particular experimental device incorporating the platform-specific parameters.
This way, we are able to distinguish clearly the experimental research towards better qubits, and the industrial and societal applications that need to be developed and executed on a quantum device.
The first case offers experimental physicists with a full-stack experimental platform using realistic qubits with decoherence and error-rates while the second case offers perfect qubits to the quantum application developer, where there is no decoherence nor error-rates.
We conclude the paper by explicitly presenting three examples of full-stack quantum accelerators, for an experimental superconducting processor, for quantum accelerated genome sequencing and for near-term generic optimisation problems based on quantum heuristic approaches.
The two later full-stack models are currently being actively researched in our group.
\end{abstract}

\section{Introduction}

\begin{figure}[hbt]
\centering
\includegraphics[width=1.0\textwidth]{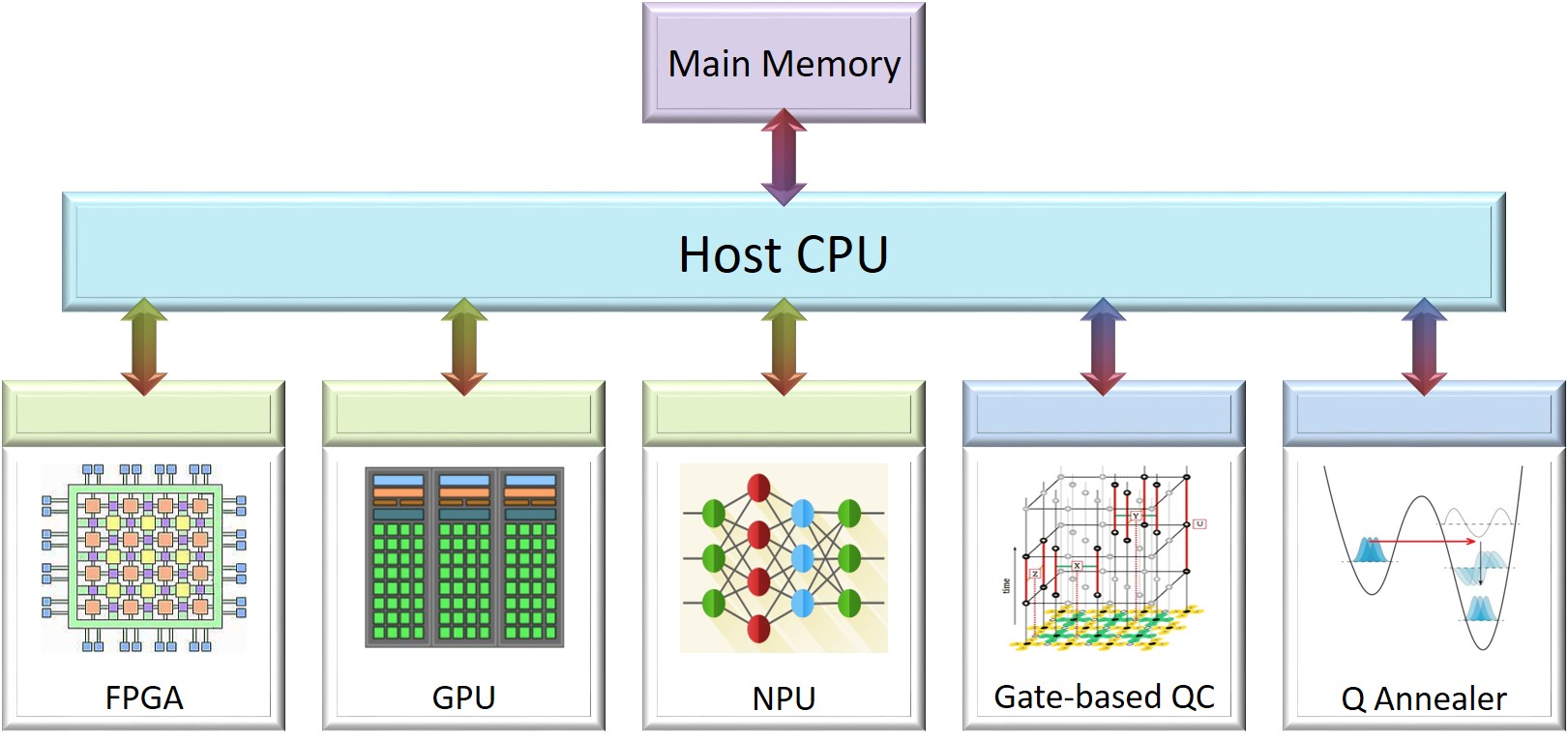}
\caption{System architecture with heterogeneous accelerators}
\label{fig:accel}
\end{figure}

The history of computer architecture dates back various decades and has been very evolving.  
An important extension is the emergence of accelerators~\cite{vassiliadis2004} as specialised processing units to which the host processor offloads suitable computational tasks.  
Recently, computer architecture research is getting more focused on quantum computing.
In the next 5 to 10 years of quantum computer development, it does not makes sense to talk about quantum computing in the sense of a universal, Turing computer that can be applied in any kind of application domain.  
Given the recent insights leading to e.g. Noisy Intermediate-Scale Quantum~(NISQ) technology as expressed in~\cite{preskill2018quantum}, 
we are much more inclined to believe that the first industry-based and societal relevant application will be a hybrid combination of a classical computer and a quantum accelerator.
It is based on the idea that any end-application contains multiple computational kernels and the properties of these parts are better executed by a particular accelerator which can be, as shown in Figure~\ref{fig:accel}, either field-programmable gate arrays~(FPGA), graphics-processing units~(GPU), neural processing units (NPU) like Google's tensor processing unit, etc. 
The formal definition of an accelerator is indeed a co-processor linked to the central processor that is capable of accelerating the execution of specific computational intensive kernels, as to speed up the overall execution according to Amdahl's law. 
We now add two classes of quantum accelerator as additional co-processors. 
The first one is based on quantum gates and the second is based on quantum annealing.
The classical host processor keeps the control over the total system and delegates the execution of certain parts to the available accelerators.
  
Computer architectures have evolved quite dramatically over the last couple of decades.
The first computers that were built did not have a clear separation between compute logic and memory.
It was only with von Neumann's idea to separate and develop these distinctly that the famous von Neumann architecture was born.
This architecture had for a long time a single processor and was driven forward by the ever increasing number of transistors on the chip, which doubled every 18 months.
In the beginning of the 21st century, the single cores became too complex and did not provide any substantial processing improvement.
This led to the incorporation of multiple cores.
The homogeneous multi-core processor dominated the processor development for a couple of years but companies such as IBM and Intel started understanding that heterogeneity is the right way forward to improve the compute power.
GPUs and FPGAs are seen as natural extensions of the computer architecture, implying that the quantum accelerator would be a logical next step.
    
\begin{figure}[htb] 
    \centering
    \captionsetup{justification=centering}
    \subfigure[Experimental full-stack with realistic qubits]
    { 
        \centering
        \includegraphics[width=.42\textwidth]{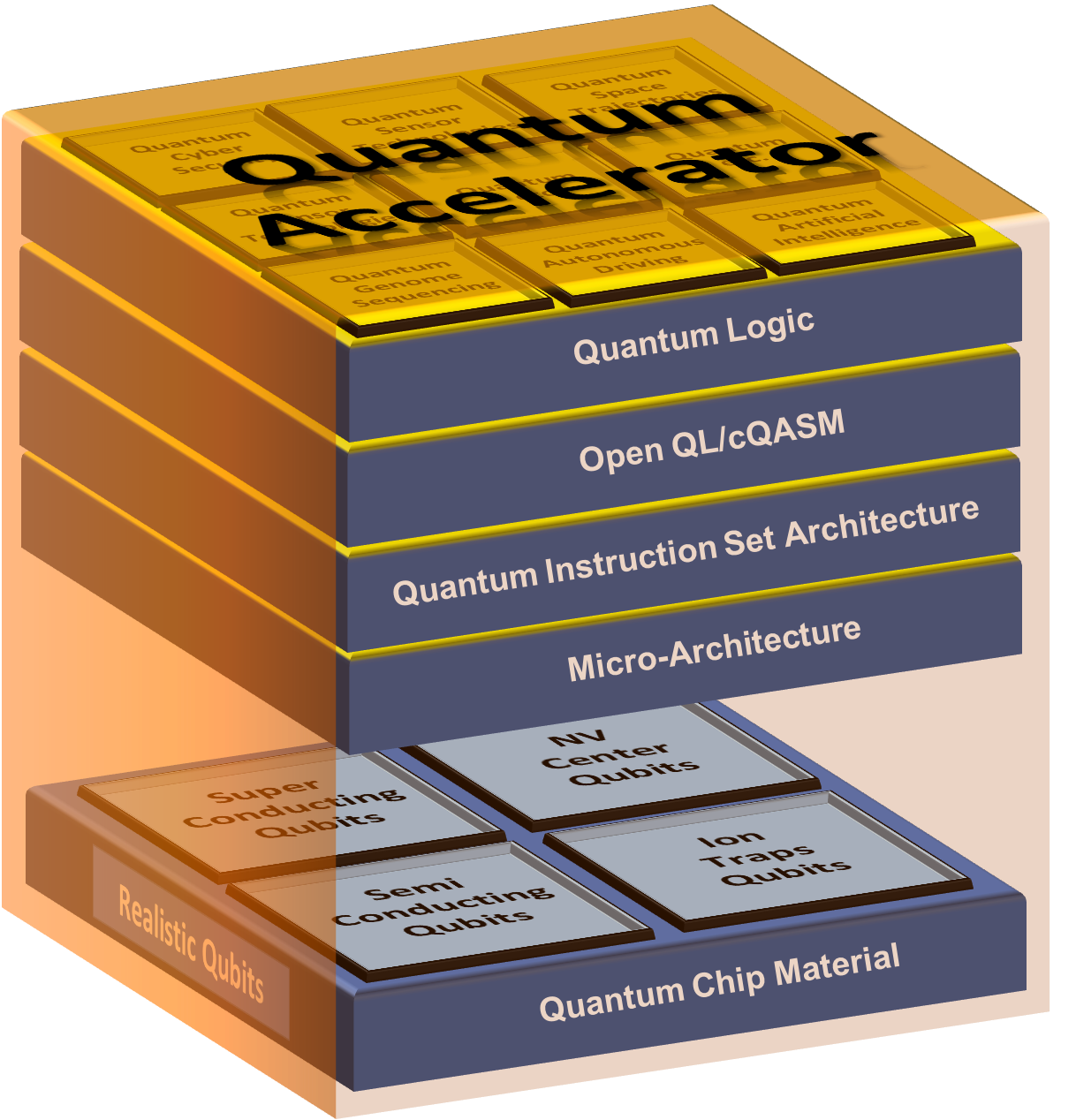}
        \label{fig:QAX} 
    } 
    \subfigure[Simulated full-stack with perfect qubits]
    { 
        \centering
        \includegraphics[width=.42\textwidth]{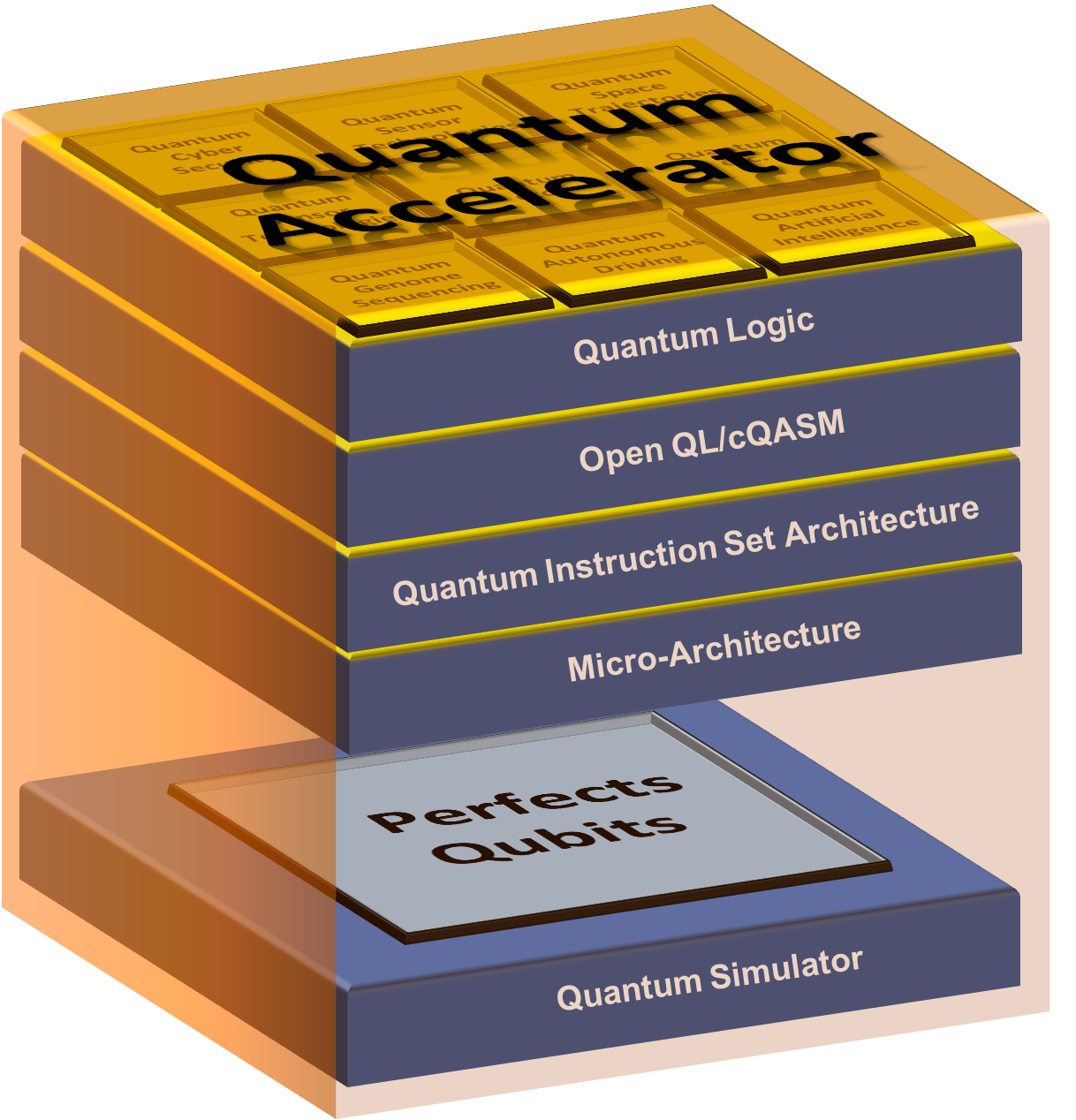}
        \label{fig:QAS}
    }
    \caption{Two approaches for full-stack quantum accelerators} 
    \label{fig:qcube}
\end{figure}

In the quantum computing world, there exist two important challenges.  
The \textbf{first} is to have enough numbers of good quality qubits in the experimental quantum processor.
The current competiting qubit technologies include ion traps, majoranas, semi-conducting and superconducting qubits, NV-centers and even graphene.  
Improving the overall status of the qubits is challenging as these suffer from decoherence that introduces errors when performing  quantum gate operation.
It is only when the quantum physical community overcomes those challenges that the quantum accelerator will be a widespread adopted solution.  
This direction is shown in the left picture  of Figure~\ref{fig:qcube} where different quantum technologies are depicted in the lowest layer.
The \textbf{second} challenge is to formulate at a high level the quantum logic that companies and other organisations need to be able to use high-performance accelerators for certain computations that can only run on the quantum device.
This requires a long-term investment in terms of people and technical know-how from companies that want to pursue this direction and reap the benefits.
The right part of Figure~\ref{fig:qcube} shows the industrial commitment to think about the required quantum logic that can be executed using the full-stack, evaluated and tested on a quantum simulator. 
It is important to emphasise that the qubits are called perfect qubits that do not decohere or have any other kind of errors generated by them.
With the emergence of huge amounts of data, commonly called big data, it is understood that this paradigm is not scalable to super-large data sets.  
The key factor is the huge amount of data that needs to be processed by multiple computing cores which is a very tough problem to solve.
The data communication between the cores is a very difficult programming problem and the data management problem is substantially slowing down the overall performance.

  
Based on our group's research since 2004~\cite{vassiliadis2004} and as shown in Figure~\ref{fig:qcube}, an important concept that we have been implementing in the quantum computing world is the implementation of a full stack for a quantum accelerator as will be described later in this paper.
The basic philosophy of any accelerator is that a full stack needs to be defined and implemented.  
The last 10 to 15 years have shown a large number of accelerators that were developed as part of any modern computer architecture.  
It always consists of the same following layers: it starts at the highest level describing the logic that needs to be mapped on the accelerator.  
Examples are video processing, security, matrix computation, etc.
These application-specific algorithms can be defined in various languages such as C++ or Fortran.
In the case of FPGAs, these algorithms are translated into VHDL or Verilog.
In the case of GPUs, the language is often formulated using mathematics or other libraries and translated by the compiler to an assembly language that can be mapped on the GPU-architecture.
Especially in the case of FPGAs, there is no standard micro-architecture on which the VHDL or Verilog can be executed. 
Such an architecture needs to be developed for every application that needs to be accelerated.
The final layer is a chip based implementation of the micro-architecture combined with the hardware accelerator blocks that are needed.

\subsection*{Background}

One of the first proposals on quantum computing was written by R.\ Feynman in 1982 \cite{feynman1982simulating} which launched a world-wide research on quantum computing focusing on important low-level challenges leading to the development of  superconducting qubits, ion trap qubits or spin-qubits.  
He formulated the use of quantum computers as an important scientific instrument to allow us to understand the quantum phenomena that quantum physics tries to understand.
The design of proof-of-concept quantum algorithms and their analysis with respect to their theoretical complexity improvements over classical algorithms has also received some attention.  
However, we still need substantial progress in either of those domains. 
Qubits with a sufficiently long coherence time combined with a true quantum killer application are still crucial achievements on which the community is working.  
These are vital to demonstrate the exponential performance increase of quantum over conventional computers \emph{in practice} and are urgently needed to convince quantum sceptics about the usefulness of quantum computing such that it can become a mainstream technology within the coming 10 to 15 years.  
However, as we will describe in this paper, we need much more before any kind of computational device can been developed, which ultimately connects the algorithmic level with the physical chip.
What is needed involves a compiler, run-time support and more importantly a micro-architecture that executes a well-defined set of quantum instructions.

An interesting and quite high-level kind of description was published in 2013~\cite{van2013blueprint}. 
The authors describe their understanding of the blueprint of a quantum computer. 
They correctly emphasised the need to look at computer engineering to better understand what the similarities and differences are between quantum and classical computing.  
As mentioned before, the most important difference is the substantially higher error rate that qubits and quantum gates ($10^{-3}$) have compared to CMOS-technology ($10^{-15}$).
Guaranteeing fault-tolerant computation can easily consume more than 90\% of the actual computational activity.  
The second difference focuses on the nearest-neighbour constraint which imposes that two-qubit gates can only be applied if the qubits reside next to each other. 
The no-cloning theorem prohibits copying quantum states. 
The way that two-qubit gates are applied requires the two qubits to be sufficiently close to each other.  
They also describe a hierarchical layered structure but rather than defining these layers in terms of more computer engineering concepts, the schema is more expressed in terms of the different, relevant fields and research domains. 
Examples are Quantum Error Correction (QEC) theory, programming languages, fault-tolerant (FT) implementation and so on. 
There are also other mechanisms with undefined time costs that are necessary to make FT-quantum computing (hopefully) efficient and performing. 
Examples are state distillation for ancilla factories and the emergence of a wide variety of defects and errors, which all impose an additional burden on the micro-architecture and the corresponding run-time management.

An older but conceptually quite similar paper was published by DiVincenzo in 2000~\cite{divincenzo2000physical}.  
This article outlines 5 criteria needed to build a quantum computer: i) a scalable physical system with well characterised qubits, ii) the ability to initialise the state of the qubits to simple fiducial state, iii) long relevant coherence times, iv) a universal set of quantum gates and v) a qubit specific measurement capability. Two additional criteria needed for quantum communication are, the ability to inter-convert stationary and flying qubits and the ability to transmit flying qubits between specified locations.
Considering currently available quantum processors, we could say that they already comply to DiVincenzo's criteria and thus we already have a quantum computer.
However, an important and missing criterion is the number of qubits that we need for any kind of reasonable application. 
Depending on the application domain, the estimates of the number of qubits goes from relatively low, such as a couple of hundreds, to several billions.  
Being less critical, we could say that the first criterion explicitly formulates the size of the system, which is still a very considerable challenge to compute in a reliable way.


The rest of the paper is structured as follows.
First, we describe the various layers like, application, algorithmic logic, programming language and OpenQL compiler, micro-architecture, mapping and simulator.
Each layer is positioned with respect to real, realistic and perfect qubits.
Three examples of quantum accelerators are presented next.
Finally, our vision on various aspects of quantum computing is discussed.
\section{The Quantum Full-Stack}

In the context of quantum accelerator development, the same full-stack approach is adopted for either perfect or realistic qubits. 
The execution can be either on an experimental quantum chip or on the QX simulator. 
The highest level starts at the end-user application for which a part of that application is developed in a quantum language, such as OpenQL.
The quantum part of any industrial or societal application can be executed on any kind of available quantum prototype.
For any quantum logic that is specified, a specific and target-related micro-architecture needs to be defined and used. 
We present the considerations for the various layers in this section.
Besides gate-based quantum computing approach, we also include the quantum annealer based system/simulator in Figure~\ref{fig:layers} as we currently investigate the components of all types of architectures currently in the market.
We first introduce here the different kinds of qubit models that we support at this state of research in the quantum computer engineering field.
The real, realistic and perfect qubits are presented here, that  can be used for either purely experimental or purely application development perspective.

 \subsection{Real, realistic and perfect qubits}

An important concept that is introduced for our line of research is the use of three kinds of qubits, namely real, realistic and perfect qubits.
In this section, we define them in detail and how these relate to each other.
 
\textbf{Real qubits:} The first qubit type is the experimental qubit, called the real qubit, which refers to experimentally realised system with challenges such as decoherence and error-rates. These features need to be substantially improved for any commercially available quantum device. The real qubits are investigated by the experimental quantum physicists community.  The goal is to improve the quality of the real qubits such that these become more easy to scale to large numbers and allow for a pragmatic micro-architectural control. This implies that there is a need to study how long the qubits can stay in a particular state and maintain their fidelity, called the coherence time. Most of the real qubits go to the ground state in a very short time (ranging from micro to milliseconds) after these are created in a particular state. Adding to that, all the quantum gates that need to be applied to the qubits generate errors. In quantum gate operations the errors and the error-rates need to be better than the current $10^{-2}$ rates. \footnote{We will limit ourselves now to quantum gates but will introduce later the quantum annealing approach.} 
There are currently many quantum technologies experimenting to produce good quality qubits for reasonable quantum computation.  The use of real qubits is very important as the physicists need to understand the dynamic and static behaviour of the qubits under different circumstances. Many large companies implement physical system for quantum computing such as IBM, Google, Rigetti, D-Wave Systems, IonQ, etc. However, the quality as well as the number of these qubits is very limited and the decoherence and error-rates as mentioned before are currently problematic for application development as these tend to influence the overall result that the quantum device is computing. 

 
\textbf{Realistic qubits:}  Realistic qubits represent the third dimension in Figure~\ref{fig:qcube} and any computer architecture needs functionality to continuously monitor the quantum system to detect and recover possible errors, as we describe here. For quite a long period, the focus has been mostly on planar surface codes as it was considered one of the most promising QEC codes for short-term implementations and for scalability concerns in the FT-era and manufacturing.  
Qubits are generally manufactured in a regular 2-D lattice connectivity with only nearest-neighbour (NN) interactions. The array comprises of two kinds of qubits, namely the data and ancilla qubits. Data qubits are used to store the quantum information for the computation, whereas ancilla qubits are helper qubits which are used to detect bit-flip and phase-flip errors by performing error syndrome measurements (ESM). This implies that after every sequence of quantum gates, the system needs to measure out its state and interpret those measurements to see if an error has been produced. Given the constraints of the coherent qubit lifetime, it implies that a very large graph needs to be processes and interpreted in real-time such that any error can be identified. Measurements themselves can be erroneous and therefore need to be repeated multiple times before a final conclusion is reached. In 2018, Preskill~\cite{preskill2018} introduced a counter-argument to this approach because surface code requires too many ancilla-qubits for logical protection.   This led to the re-initiation of the small-codes which were first defined almost 20 years ago. The impact on the system architectural and compiler level is yet unclear but this is currently the focus of a lot of research. 

 \textbf{Perfect qubits:} Companies, governments and other organisations interested in building a quantum accelerator need to evaluate the availability of quantum computing resources in terms of quantum algorithms and have a way to test the correctness of the quantum logic. To serve these needs, we use perfect qubit, such that any of the erroneous behaviour arising due to qubit quality can be avoided during application development phase. These qubit modelled in the simulator do not decohere and stay in ideal state required for the algorithm. Using these perfect qubits guarantees that the end-users can verify and check the algorithm that they are working on and test if the computed results have a meaning that can be easily interpreted. We are not the only ones who use this but it is a very clear concept that separates the two directions that we are investigating in the Quantum Computer Architecture lab. As explained above, we introduce in OpenQL, a datatype which represents the perfect qubit which has a more stable behaviour than the realistic qubits. Whether or not the nearest-neighbour constraint applies, is a discretion of the designer. The compiler may or may not compute a route for the qubits. These decisions are based on the requirement and maturity of the application development stage before translating to realistic experimental testing.
 


\subsection{Industrial and societal quantum application logic}

The highest layer in the full-stack focuses on the application that needs to be developed for any organisation. On current, modern architectures, there are a large number of initiatives developed that run on either the FPGA, the GPU or the TPU as the accelerator platform. When envisioning the quantum accelerator idea, many similar topics are well suited, such as security, artificial intelligence, autonomous driving, genome sequencing, sensors and trajectories for aeroplanes and rockets. For the three application examples that we are currently developing, we assume the use of perfect qubits such that the focus can be completely given to the algorithm logic and the micro-architecture design.
\begin{enumerate}
    \item We research algorithms for accelerating quantum genome sequencing. These are motivated by the application of gene therapy and personalised medication for every single individual on earth. The treatment will be based on every person's DNA-profile that has to be generated by extensive computational processing of the reads from sequencing devices.
    \item The other example that we will discuss in this paper is for optimisation problems pervasive in operations research based on the travelling salesman problem. It is expressed as a quadratic unconstrained binary optimisation problem and can be solved both on the gate based model or the annealing model.
    \item We are also working on a quantum accelerator model in collaboration with 
    our research partners in the automotive industry focusing on autonomous and electrical cars. For confidentiality agreements, we do not go into any detail of this project.
    
\end{enumerate}

Given the potential of quantum acceleration, this top-down approach is necessary to understand how investing in the development of quantum computing has the potential to become a world-wide technology that can be used by every country, organisation or individual. 
In section \ref{3stack}, we will present the three accelerators in more detail.

\subsection{Quantum logic}

For this section, we always consider perfect qubits.
The highest level is the application layer where a potential end-user of the quantum compute power instructs what exactly needs to be computed.
Quantum computing promises to become a computational game changer, allowing the calculation of various algorithms much faster (in some cases exponentially faster) than their classical counterparts.
Especially, applications requiring manipulation of a large set of data items to produce a statistical answer are very suitable to be processed by quantum computers, which we call in this paper quantum accelerators.
Currently, there is no generally acknowledged or accepted functional domain where quantum technology would be the game changer.
Potential promising domains include physical system simulation, cryptography and machine learning.
Evidently, the cryptography domain is a clear candidate as algorithms such as Shor's factorisation showed that potentially a quantum computer can break any RSA-based encryption, as it leads to finding the prime factors of the public key~\cite{shor1994algorithms} based on which the private key can be easily calculated.
However, the cryptography domain has actively establishing a new research theme, namely the post-quantum cryptography such that the attacks emerging from such a compute power can be retaliated.

\begin{figure}[bt]
\centering
\includegraphics[width=0.7\textwidth]{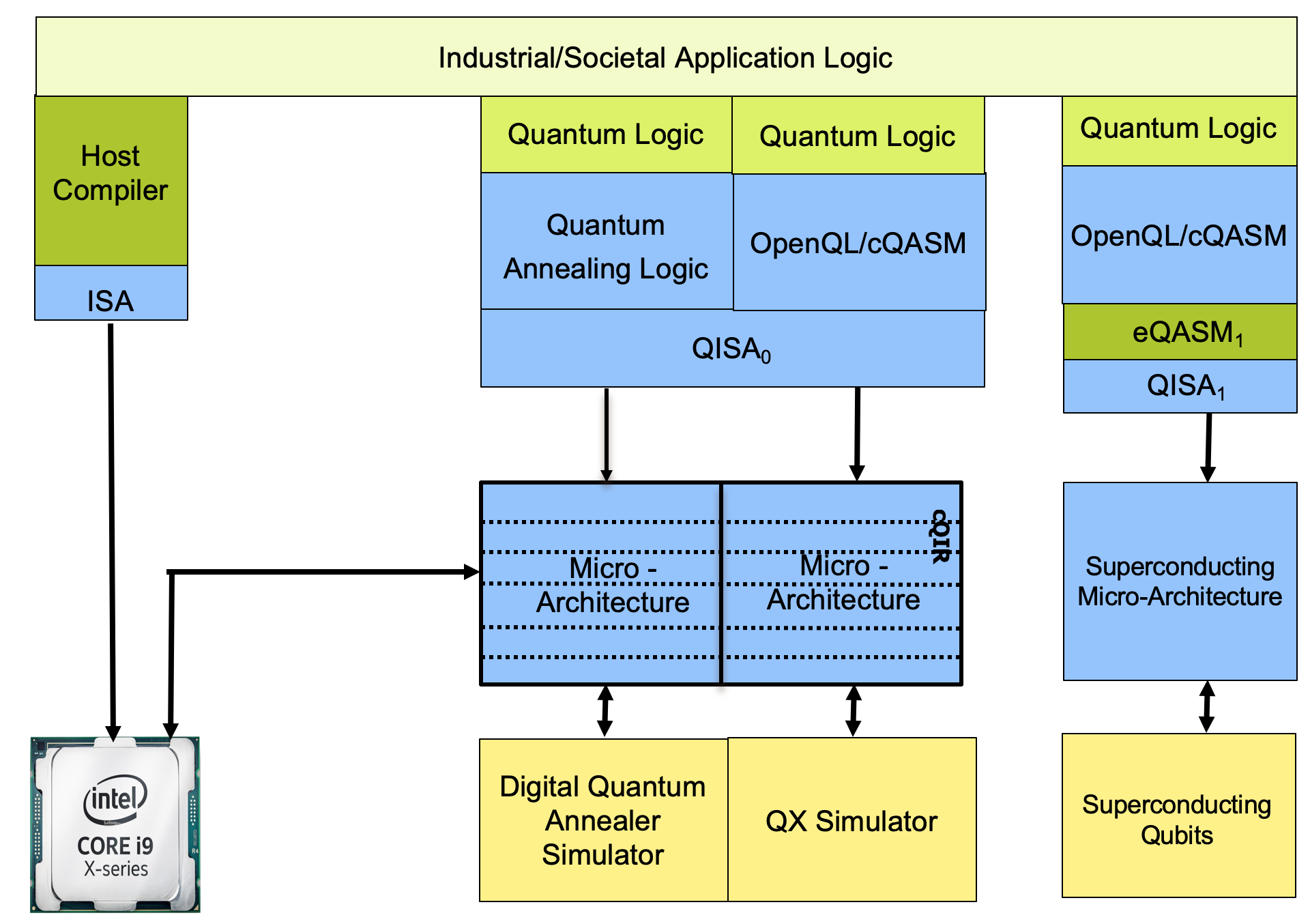}
\caption{Full-stack execution}
\label{fig:layers}
\end{figure}

Another potential application area is the biological domain where chemistry, medication and pharmacology belong to.
We focus on one such candidate application of genome sequence reconstruction.
For instance, quantum computational power would be imperative if we aim want to compute the DNA-profile of every human being in the world, which takes around one week on a large network of very powerful servers for one person's DNA.
With the availability of enough qubit capacity, the entire parallel input data-set can be evolved simultaneously as a superposition of a wave function.\footnote{By our estimate, given the size of the human genome and currently available sequencers, the number of qubits required will be around 150 logical qubits.}
This particular property makes it possible to perform the computation of the entire data-set in parallel.  
This kind of computational acceleration provides a promising approach to address the computational challenges of DNA analysis algorithms. 
The essence of accelerating sequence reconstruction is the ability to run parallel search operations on the short reads obtained from sequencing an individual DNA from a sequencing machine, onto an already available reference of the organism. 
In recent years, GPU, FPGA and cluster computing frameworks like Hadoop and Spark have been used to reduce the total run-time. 
Potentially, quantum computation offers a fundamentally different way to address the enormous volume of data by employing superposition of reads in the search process, thereby reducing the memory requirement maybe even exponentially.
The quantum search primitive (Grover's search) itself is provably optimal~\cite{zalka1999grover} over any other classical or quantum unstructured search algorithm.
The rather modest quadratic speedup in cycles however becomes extremely relevant for industrial application due to the total CPU run-time involved in the big data manipulation (in order of 1000s of CPU hours~\cite{houtgast2018hardware} for a single human DNA sequence reconstruction).


\subsection{Programming language, compiler and run-time support}

The quantum algorithms and applications presented in the previous section can be described using a high-level programming language such as Q\#~\cite{svore2018q}, Scaffold~\cite{ abhari2012scaffold}, Quipper~\cite{ green2013introduction} or OpenQL~\cite{khammassi18} and compiled into a series of instructions that belong to the (quantum) instruction set architecture.

\begin{figure}[bt]
\centering
\includegraphics[width=0.75\textwidth]{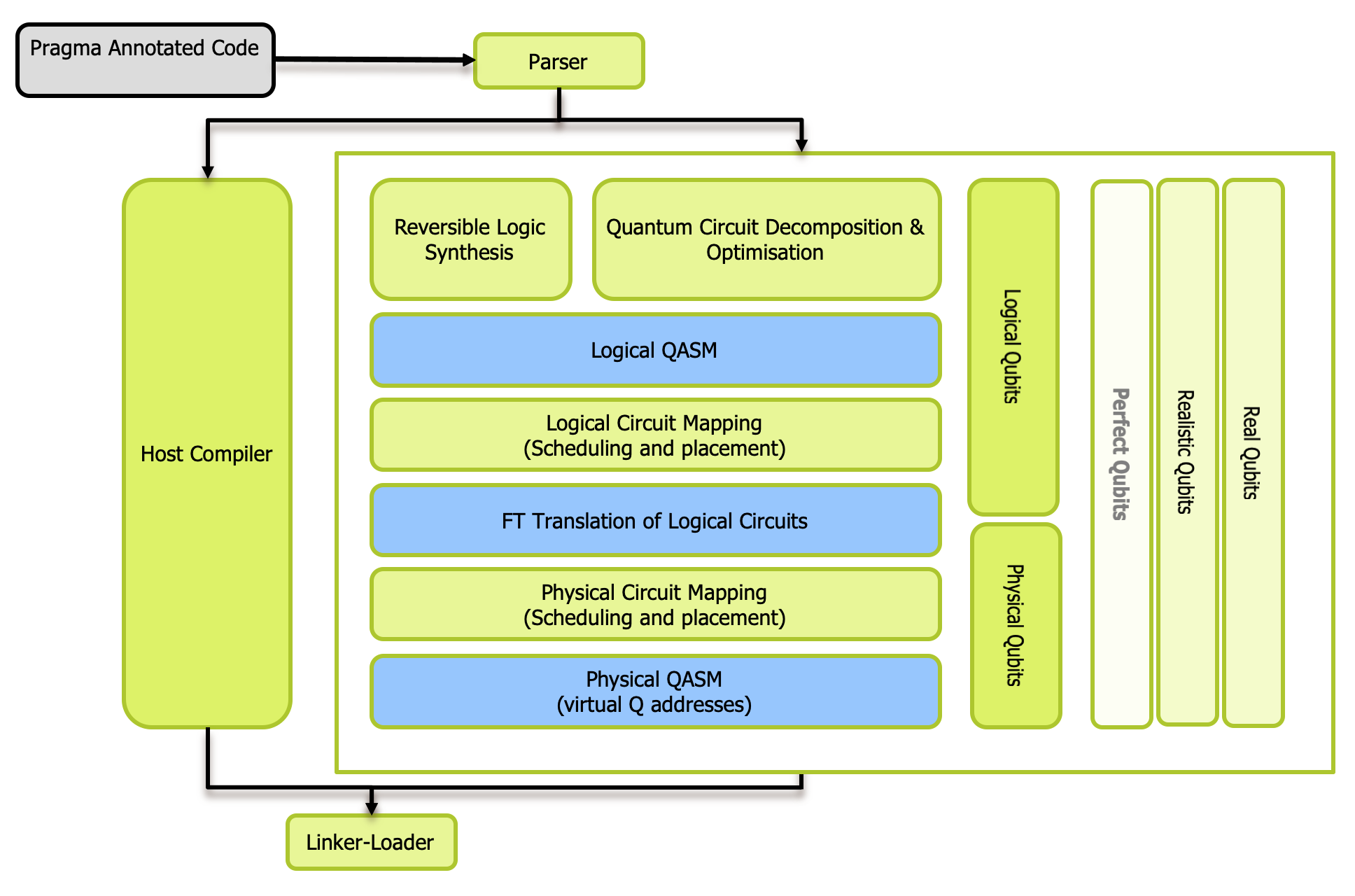}
\caption{Compiler infrastructure}
\label{fig:compiler}
\end{figure}
 
Consistent with our distinction between perfect, realistic and real qubits, the compiler is capable of adapting to the requirements of the end-user.
So there is an option that translates the qubits in perfect, realistic or real manner.
As shown in Figure~\ref{fig:compiler}, the compiler infrastructure for such a heterogeneous system consists of the classical compiler for the host processor combined with the quantum compiler.
It is important to note that the architectural heterogeneity where classical processors are combined with different accelerators such as the quantum accelerator, imposes a specific compiler structure where each compiler part can target the different instruction sets and ultimately generates one binary file which can be executed on different instruction set architectures.
For the computer architecture envisioned in our research, any high-level implementation of the system application will consist of two interleaved types of logic: the classical logic which will be executed by the micro-architecture of the controlling processor and the quantum logic which will be mapped onto the quantum processor.
The quantum logic can be encapsulated by classical language structures such as decision and loop constructs.
The micro-architecture extracts the quantum part and send it to the quantum processor.
 
As we adopt the quantum circuit model as a computational model, the quantum compiler translates the quantum logic into quantum circuits for which reversible circuit design, quantum gate decomposition and circuit mapping are needed.
The output of this compiler is a series of instructions, expressed in a quantum assembly language, such as cQASM, that belongs to the defined instruction set architecture. \footnote{QASM is one candidate for such a language and was originally produced by Nielsen and Chuang to generate the \LaTeX~figures for the quantum circuits for their book.} The definition of a shared quantum assembly language is a key challenge such that there is uniformity in the algorithmic descriptions of different research groups. 
 
 \begin{enumerate}
 
     \item \textbf{Real qubits:} The OpenQL compiler can  generate code that physicists can use for testing the behaviour of the qubits, taking all kinds of errors and decoherence into account. An important exercise is to examine the fault-tolerance (FT) of the quantum circuits. A central issue for any quantum technology is its fragility, implying that the qubit superposition state disappears quite rapidly. First, the coherence time of real qubits is extremely short. For example, superconducting qubits may loose their information in tens of microseconds~\cite{DiCarlo2015, corcoles2015demonstration}. Second, quantum operations are unreliable with error rates around $0.1\%$~\cite{Martinis2015}. As mentioned above, in January 2018, Preskill \cite{preskill2018} emphasises that early stage quantum computers should be based on Noisy Intermediate-Scale Quantum (NISQ) technology with much less ancilla qubits for Quantum Error Correction (QEC) activities. 
     Quantum Error Correction is more challenging than classical error correction, due to the no-cloning theorem, which states that (unknown) quantum states cannot be copied. This makes the classical way of creating several copies of the same bit impossible. In addition, quantum errors are continuous and any measurement will destroy the information stored in qubits. The basic idea of QEC techniques is to use several physical imperfect qubits to compose more reliable units called \textit{logical qubits} based on a specific quantum error correction code~\cite{lidar, shor1995scheme, steane1996multiple, calderbank1996good, gottesman1996class, bombin2006topological, fowler2012surface}. This is what scientists looking at physical implementations of qubits have been doing such that it is relatively simple to generate and test a super- or semiconducting qubit.
     
     \item \textbf{Realistic qubits:}  Similar to real qubits, it is also possible to simulate the behaviour of realistic qubits such that we have a better understanding of the impact of realistic error models, better error-rates and longer coherence times on the overall quantum circuit performance, the micro-architecture needed to control them and so on. Therefore, there is the option to compile for realistic qubits such that the duration of a quantum gate operation is shorter or less error-prone. It can also lead to better investigation of the qubit plane topological constraint and the associated routing algorithm required for multi-qubit gate operations.
     
     \item \textbf{Perfect qubits:} The compiler can also target the use of perfect qubits. As defined above, that implies that these qubits live as long as they are needed and have principally no error-rates in the quantum gates that are executed. Depending on the state of the execution platform, connectivity constraints can be imposed for mapping and routing. When we generate everything in terms of perfect qubits, that also implies that there is no separation anymore between logical and physical qubits as there is no requirement for error coding.  
\end{enumerate}

\subsection{Quantum micro-architecture}
Any computer has a series of instructions which can be executed on the dominant processor.
To this purpose, any kind of processor has a particular architecture capable of executing any sequence of the legitimate instructions.
This also holds for the quantum processor, which also has a series of instructions that it can execute, some of which are classical logic and others are the quantum instructions that will be executed on the quantum chip.
So the quantum accelerator will consist of two components: the classical and digital micro-architecture part that has a classical processor to execute part of the accelerator logic and the quantum chip that contains the qubits that need to be executed in an analogue way.


\begin{figure}[bt]
\centering
\includegraphics[width=1.0\textwidth]{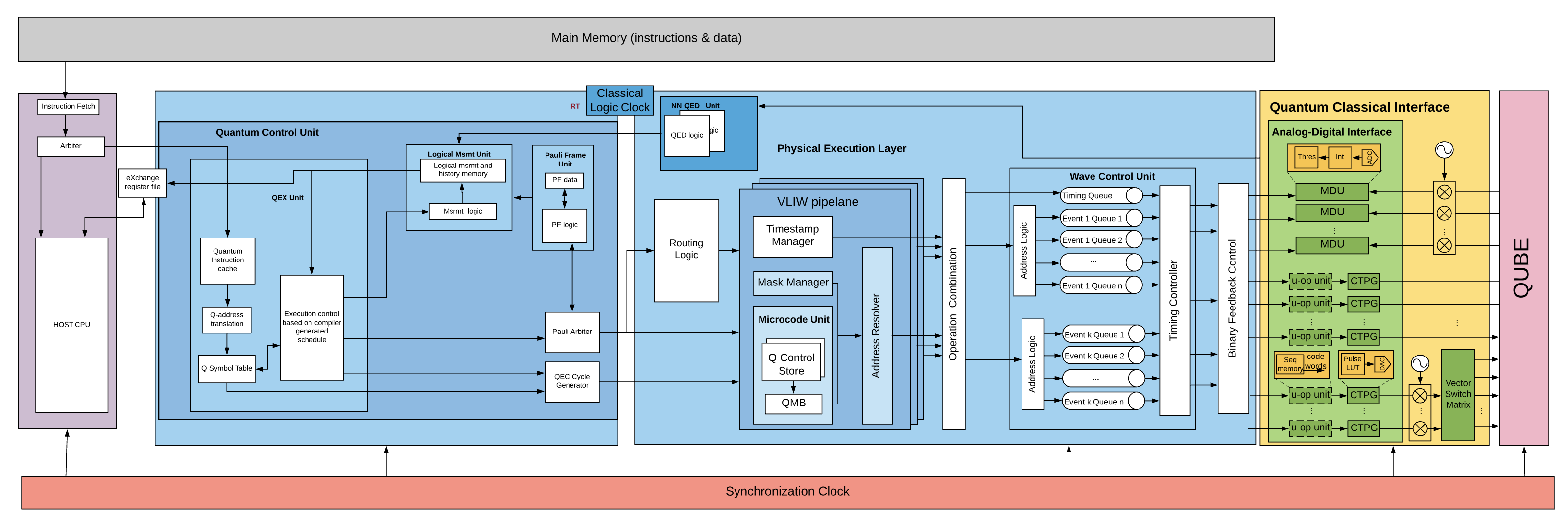}
\caption{\label{fig:microarchitecture}An Example of a general-purpose Micro-Architecture}
\end{figure}

Essential to any kind of computational device is the presence of one or multiple computer architectures that are responsible for executing the instructions that are delegated to the co-processor.
The architecture of a machine connects the physical hardware to the applications that can run (on that hardware) and dictates how instructions are executed. 
This is also true for the case of a quantum accelerator.
For the quantum algorithms to be understood by the quantum accelerator, a low level representation of the quantum instructions is required that the classical control hardware of the quantum chip can understand.
This is known as the Quantum Instruction Set Architecture (QISA).
The content of the QISA can be modified for each accelerator logic that needs to be implemented.
One example of a micro-architecture is given in Figure~\ref{fig:microarchitecture}.
For any micro-architecture, there are a number of properties that we have to estimate, such as the appropriate instruction-length, pipeline depth (for parallel quantum gates) and targeting multiple control channels per single instruction. 
Based on these principles, the basic blocks are constructed, such as timing control unit and the microcode instruction set of the overall micro-architecture.

\begin{enumerate}
    \item \textbf{Real and realistic qubits:} To accommodate quantum processor development, we look at the experimental algorithms that the physics community are interested in, such as randomised (single and double) qubit gates. This phase would also comprise of hardware assessment and characterisation to meet the timing-precision and signal synchronisation requirements for a specific qubit-technology. In a later phase, the experimental implementation will need to include error-correcting codes in the pipeline. A system-on-chip running a quantum error-decoder would enable faster development and debugging capabilities for QEC on hardware. Area utilisation and power consumption of such a firmware would become a necessary consideration at this point, depending on the size of decoders. The development and testing of this platform would be done on both the QX simulator and the physical quantum processing unit. Mapping of the quantum circuit also needs to be addressed as part of the compilation process. 

    \item \textbf{Perfect qubits:} We do not yet have a full implementation of the micro-architecture for logic expressed in terms of the perfect qubits. Later in this paper, we present a tentative micro-architecture for quantum genome sequencing (QGS), which is one of the accelerators that we are working on. It is important to define the QISA needed for QGS and fine-tune the corresponding micro-architectural blocks needed to execute the quantum instructions on the QX simulator.

\end{enumerate}

\subsection{Mapping of quantum circuits}
Mapping of quantum circuits is considered in two different contexts: the first is when applied on small real quantum processors and the second one targets a simulation engine that addresses larger number of qubits. Depending on the test objective, we can either take into account large number of qubits or stay at a small scale and closer to the experimental state-of-the-art.

\begin{enumerate}
 
    \item \textbf{Real qubits:} When targeting a real quantum processor, the mapping of circuits is an important topic as described in~\cite{lin2015paqcs,dousti12min}. The circuit description of the algorithms does not usually consider a physical location of the qubits and assumes that any kind of interaction between qubits is possible. However, real qubits need to be placed on a specific physical qubit layout that will limit the possible interactions between these, leading to an increase of the circuit latency. It is therefore important to optimise the mapping process that includes the following:
    %
    \begin{itemize}[nolistsep,noitemsep]
        \item \textbf{Scheduling of operations:} The parallelism of current quantum algorithms is pretty limited but applying classical scheduling methods and techniques, the inherent parallelism of the logical qubits can be exploited. Depending on the chosen QEC, different constraints apply to the scheduling problem. For instance, in defect-based surface codes (SC), single-control multi-target CNOT gates are possible whereas planar-based surface only supports single-control single-target CNOT gates. Furthermore, other limitations such as the number of available frequencies to control the qubits can also affect the scheduling process and restrict the parallelism.    
        \item \textbf{Placement and routing of qubits:} As mentioned before, most of the current quantum technologies are pursuing a 2-D array of qubits with only NN-interactions. This means that 2-qubit (physical) operations are only possible between adjacent qubits. It also impacts the placement of logical qubits. For instance, a CNOT between two planar-based SC qubits can theoretically be performed transversally, i.e. applying pairwise CNOT gates to each pair of data qubits in the sub-lattices. However, it is not possible to implement such a transversal gate in a 2-D array requiring techniques such as lattice surgery~\cite{horsman2012surface} where planar-based SC qubits still need to be placed next to each other. 
        Finally, not all qubits can be placed in the necessary adjacent positions. Therefore, some of them will have to be moved or routed for which the compiler will insert a MOVE-operation for the run-time routing logic.
    \end{itemize}
    
    \item \textbf{Realistic qubits:} This is the quantum processor where the qubits are not yet fully realised in any experimentally designed qubit processor. Realistic qubits imply that we are focusing on experimental processors but have modified some parameters in the overall design to understand the impact, for instance, a different topology, a different error distribution, the number of qubits, etc.
    \begin{itemize}[nolistsep,noitemsep]
        \item \textbf{Scheduling of operations:} Assuming that we also have parallelism between the qubits when executing a quantum circuit, we have to understand what the scheduling is of qubits and, for instance, how CNOT-gates need to be implemented to have a successful execution of the quantum gates. Comparison between the behaviour of defect-based SC qubits single-control multi-target CNOT gates and planar-based surface that only supports single-control single-target CNOT gates needs to be investigated.
        \item \textbf{Placement and routing of qubits:} Even with realistic qubits, we still have the challenge to take the NN-interaction constraints into account.  Preskill's paper and talk beings to our awareness the limitations on the experimental physicists. Small codes maybe be more relevant in this regime.
        Similar to real qubits, it also impacts the placement of logical qubits. It is important to understand if we have similar constraints as the real qubits when we are using the realistic qubit paradigm. We also need to understand if we need a similar MOVE-instruction to put the qubits close to each other.
    \end{itemize}
    
    \item \textbf{Perfect qubits:} When the algorithmic behaviour and content is not yet defined, which is the case in most of the situations, it is important to be able to use perfect qubits that are more reliable and predictable than the experimental ones, as that have no decoherence and execute reliably the quantum gates of the quantum circuit.
    \begin{itemize}[nolistsep,noitemsep]
        \item \textbf{Scheduling of operations:} With perfect qubits, we have the freedom to impose or relax similar kind of restrictive scheduling instructions on their behaviour. 
        \item \textbf{Placement and routing of qubits:} Also for this feature, it depends on how much freedom the algorithm designers needs to experiment with the algorithm they are designing. The more restrictive we are in the placement and routing, the more difficult is becomes. In a more relaxed situation, the designer enjoys more possibilities to experiment and test the algorithm.
    \end{itemize}

\end{enumerate}

\subsection{QX simulator}

The QX simulator, as shown in Figure~\ref{fig:layers} was developed in our group as a platform to simulate quantum operations on either realistic or perfect qubits.
The QX engine can execute any quantum logic expressed in OpenQL and translated by the compiler to cQASM, the common quantum assembly language.
The assumed micro-architectural layer encapsulating the QX simulator executes the cQASM instructions by sending the quantum instruction to QX, which then executes it, measures the qubit states and sends back the results to the micro-architecture.
The QX simulator is scalable based on the underlying host processor, and is capable of simulating with up to 35 fully-entangled qubits on a laptop PC, which are either perfect or realistic.
The main advantage of a platform like QX is to provide application developers, computer scientists and computer engineers the tools to model and test designs before experimental implementation on quantum processors.
A order of 50 fully entangled qubits already give a lot of possibilities to test the application in a proof-of-concept simulation.
We can also use the different kinds of qubits that we presented in this paper.
 
 \begin{enumerate}
     
     \item \textbf{Realistic qubits:} Whenever we are interested in running quantum circuits on real hardware, we need to be able to introduce error models for the qubit or gate operations, at the simulation level of realistic qubits. Current quantum error rates do not go beyond $10^{-2}$ so there is a need to understand the impact of error rates in the order of $10^{-5}/10^{-6}$. The errors will have an effect on the real qubits as well as the quantum gates. Using the QX simulator on such realistic qubits, we can investigate beyond simplistic error models such as the depolarising model (where every quantum gate is followed by some error, drawn from a uniform distribution of the different errors than can follow the Pauli gates X, Y or Z.) It can be extended to other error distributions which are more realistic sketching the extensions that the quantum physics research community need to address.
     
     \item \textbf{Perfect qubits:} For application development, there is the need to execute the quantum logic to verify the computed results of the algorithm in the functional sense.  The QX simulator is capable of assuming the non-emergence of errors. The current stage of research on quantum genome sequencing algorithm uses the QX simulator in this mode of development. In principle, any universal quantum logic can be executed on the simulator, the result can be measured and fed back to the micro-architecture.

 \end{enumerate}
 

\section{Three Full-Stack Architecture Examples} \label{3stack}

In this section, we present and briefly describe three implementations of the full-stack.
The first was developed for the experimental design of superconducting qubits, the second is being implemented for accelerating genome sequencing on quantum logic, while the third application involves optimisation problems.
The full-stack as shown in Figure~\ref{fig:qcube} is used as the basic structure.

\subsection{Full-stack for real, super-/semi-conducting qubits}

Here we present the developed micro-architecture for the superconducting quantum chip based on an experimental implementation of all the components that were defined and needed for the quantum research collaborations in our department.
The end-to-end pipeline involves writing an algorithm, up to sending the analogue pulses to the qubits.
It starts with a high level quantum algorithm which is useful for the physicists. 
We have been focusing on randomised bench-marking experiments for one or two qubits which was written in OpenQL.

The code is translated by the OpenQL compiler into our version of the Quantum Assembly language, cQASM. 
As a logical extension of cQASM, the compiler then translates that version to an executable QASM, called eQASM, which supports, in principle for any quantum technology, taking low-level information into account, such as gate times, topology etc.
It basically means that there is a second back-end compiler pass that translates cQASM into the eQASM version.

\begin{figure}[hbt]
\centering
\includegraphics[width=0.8\textwidth]{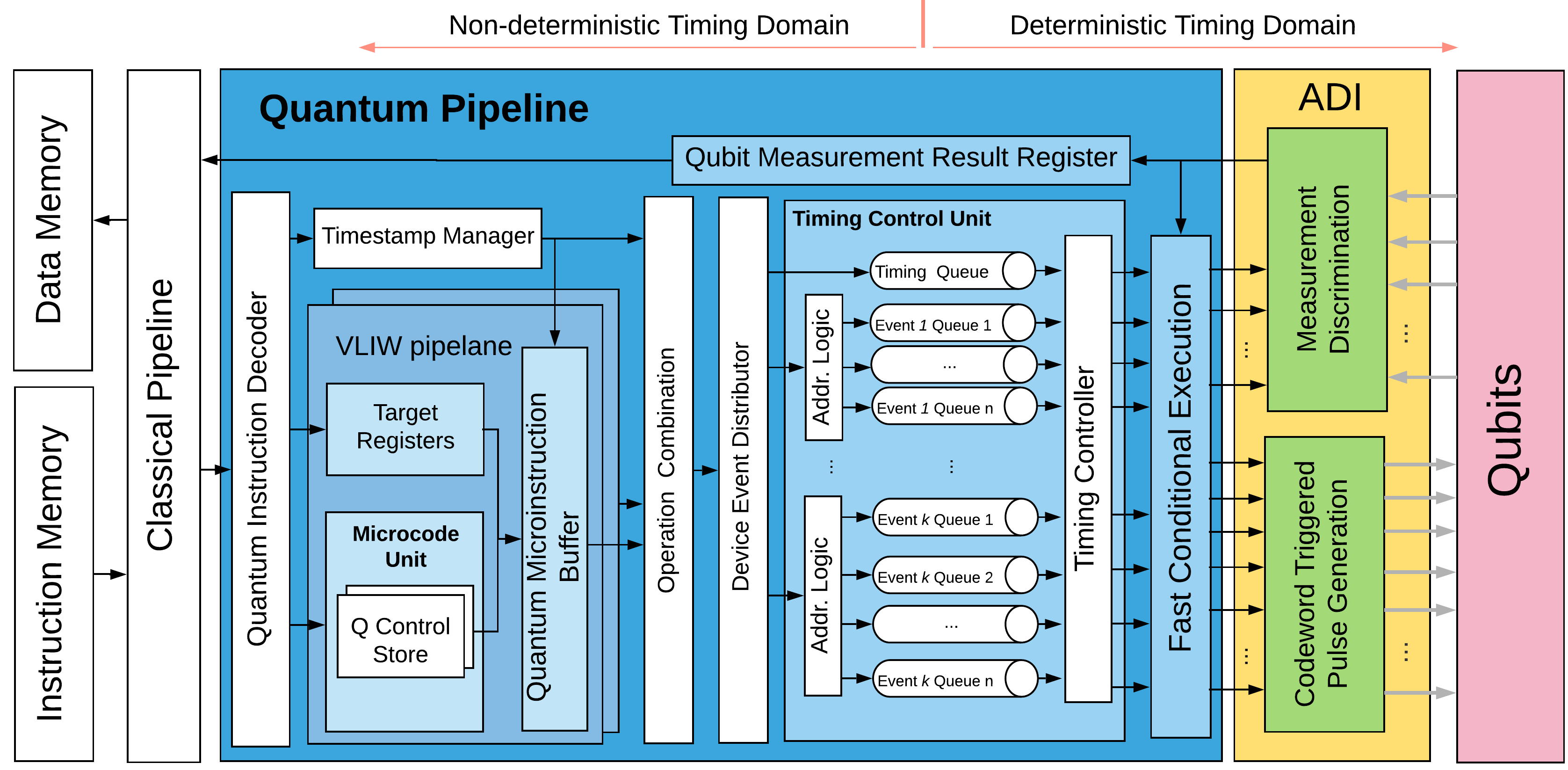}
\caption{Experimental implementation of the micro-architecture for super-conducting (real) qubits}
\label{fig:qumav2}
\end{figure}

    

Based on the cQASM code, the compiler generates the eQASM instructions which can be executed by the micro-architecture, as shown in Figure~\ref{fig:qumav2}~\cite{fu2018eqasm}. 
The eQASM is then executed and at run-time translated into the horizontal micro-code version which ultimately sends the micro-operations to the queues.
\footnote{Described in a recently submitted paper to Arxiv: X. Fu et al., eQASM: An Executable Quantum Instruction Set Architecture.}  
From that level on, the timing execution requirements are very strict and need to be precise up to the nanosecond level.
The code-words that are generated by the micro-code unit will ultimately be translated in an analogue pulse and sent to the qubit chip.

This micro-architectural demonstration was done for two quite different quantum technologies: one for the superconducting qubit chip and one for the semiconducting quantum chip.
The specific combination of the micro-architecture design parameters, the c/eQASM compiler passes and the micro-code unit proved very useful. 
Especially the last two options allowed us to re-target the same micro-architecture to two different quantum technologies and the only changes that were needed are the configuration file for the compiler and the implementation of the micro-code unit needed for the specific quantum technology to make sure the analogue pulses, stored in the analogue-digital interface (ADI), were available.

\subsection{Full-stack for quantum genome sequencing on perfect qubits}

Genome sequencing involves taking fragments of the DNA (called, short reads) from the sequencing machines and stitching them together to reconstruct the original genome of the individual. 
Reconstruction can either be carried out by aligning these reads to an already available reference genome, or in a de novo assembly manner. 
This requires the algorithmic primitive of searching an unstructured database or graph-based combinatorial optimisation respectively. 
Translating such quantum kernels to an efficient implementation on a quantum accelerator requires in-depth tuning of both an architecture-aware quantum algorithm and the underlying micro-architecture.

\begin{figure}[hbt]
\centering
\includegraphics[width=\textwidth]{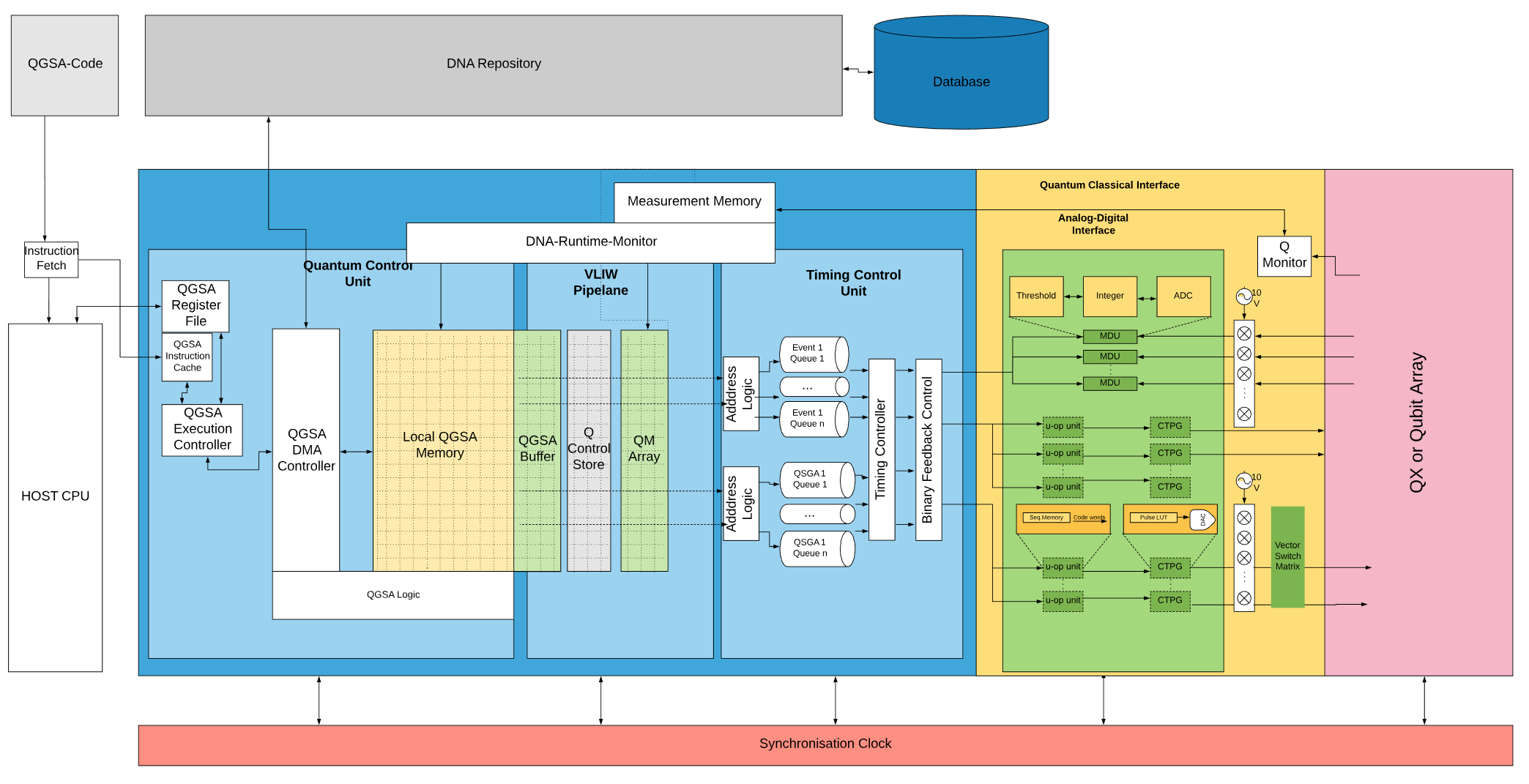}
\caption{A new micro-architecture for the quantum genome sequencing accelerator}
\label{fig:qmicrodna}
\end{figure}

We have obtained initial results from combining domain-specific modification on the Grover's search~\cite{grover1997quantum} and quantum associative memory~\cite{ventura} approaches.
This new alignment algorithm, described and analysed in~\cite{AritraMSc}, has been tested on the QX simulator platform.
The reference DNA is sliced and stored as indexed entries in a superposed quantum database giving exponential increase in capacity.
The designed algorithm~\cite{sarkar2019algorithm} considers inherent read errors in the sequence, incorporating the requirement for approximate optimal matching.
A quantum search on the database amplifies the measurement probability of the nearest match to the query and thereby of the corresponding index.
Due to the reference database and index, being entangled, the closest-match index can be estimated.
Current explorations involves designing optimisation algorithm for genomics applications using near-term Quantum Machine Learning~(QML) primitives like the Quantum Approximate Optimisation Algorithm~(QAOA).


As already mentioned, the proposed  quantum accelerator will not be a standalone machine, but rather a quantum co-processor that will be part of a heterogeneous system in which classical processors are connected to the quantum accelerator.
Each processor will have its own instruction set.
A first tentative view of the quantum genome sequencing micro-architecture is shown in Figure~\ref{fig:qmicrodna}.  



There is need for run-time support to coordinate the activities of the different micro-architectural components and, as discussed, be responsible for the run-time routing of qubit states for two-qubit gates. 
In the quantum accelerator, the executed instructions generally flow through modules from left to right. 
The pink block on the right of the figure represents the QX simulation platform or an implementation of a quantum chip on which the test-runs of the quantum genome sequencing algorithms will be performed. 
The rest of the large (blue) block represents the micro-architecture. 
The DNA data-sets is to be retrieved from an external classical database and transported to a local memory in the quantum accelerator. 
The size of the local memory will depend on the capabilities of the QX simulator platform and how that information is encoded. 
This research is based on the large-scale micro-architecture simulation platform that we have already developed.
Using the QX simulator platform makes it possible to rapidly develop hardware prototypes and verify their behaviour and performance before a FPGA implementation is started. 
The set of queues will be relevant for feeding the DNA information to the qubit chip and for defining how the quantum gates are applied.  


In a specific qubit plane topology, qubits will have to move around so that two-qubit gates can be applied on adjacent qubits.
It is a prevailing idea that quantum compilers generate technology-dependent instructions~\cite{svore2006layered, abhari2012scaffold, haner2016software}. 
However, not all technology-dependent information can be determined at compile time, because some information is only available at run-time due to hardware limitations, for instance qubits that need to be re-calibrated. 

For testing the functionality of the algorithm, we use artificial DNA sequences that preserve the statistical and entropic complexity of the base pairs in biological genomes; yet in a reduced size so that they can be efficiently simulated in a classical architecture with qubit limitations.
This implies understanding which run-time and thus routing support will be necessary to make sure that the quantum accelerator always has enough data to process and that they are in adjacent positions when necessary.

From an algorithmic perspective, near-term quantum optimisation algorithms employ the variational principle, where a shallow parameterised quantum circuit is iterated multiple times while the parameters are optimised by a classical optimiser in the Host-CPU.
This model of Hybrid Quantum-Classical (HQC) algorithms requires fast feedback between the quantum accelerator and the  real-time circuit/instruction generator (i.e. the compiler and the micro-architecture).
Since most quantum algorithms expect a statistical central tendency over multiple measurement, the expected probability of the solution state can be calculated inside the quantum accelerator itself, aggregating the measurements over multiple runs.

\subsection{Full-stack for quantum optimisation on hybrid quantum accelerators}

\begin{figure}[htb] 
    \centering
    \subfigure[Quantum Accelerator Model]
    { 
        \includegraphics[width=.36\textwidth]{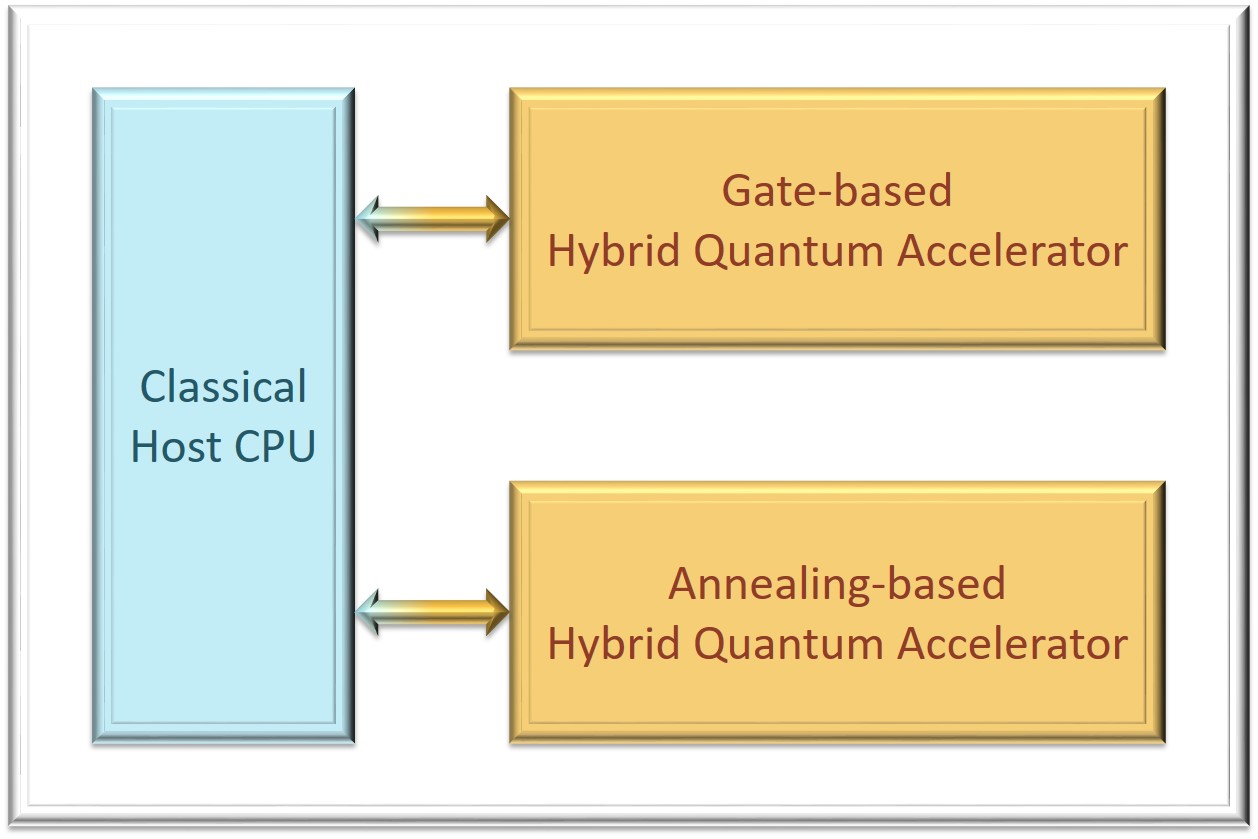}
        \label{f_hqama} 
    } 
    \subfigure[Hybrid Quantum Accelerators]
    { 
        \includegraphics[width=.36\textwidth]{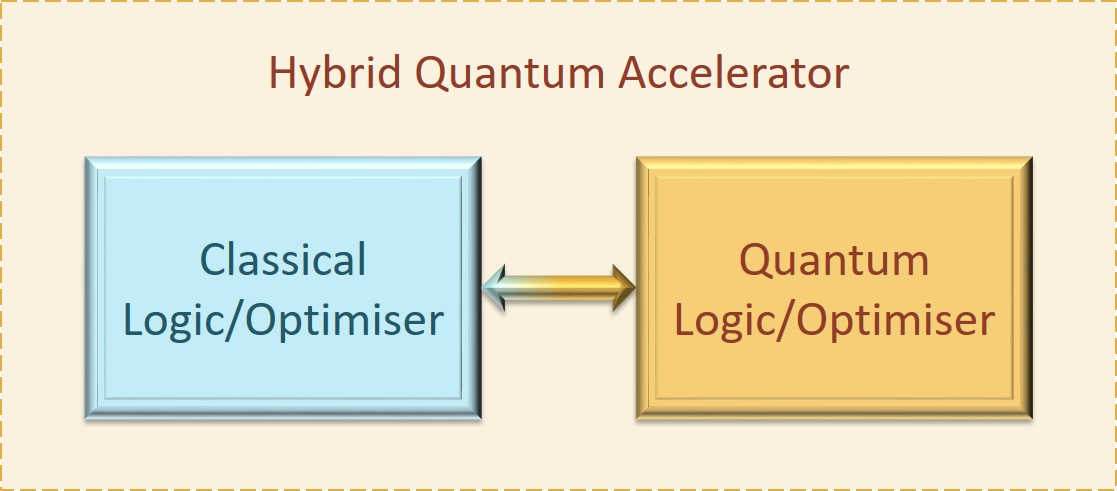}
        \label{f_hqa} 
    }
    \caption{Model for near-term quantum-accelerated optimisation} 
\end{figure}

Optimisation problems are ubiquitous and well-suited for near-term quantum acceleration.
In this stack model a generic execution model for optimisation problems is considered as shown in Figure~\ref{f_hqama}.
Near-term quantum processors will be limited in size (number of qubits), quality (noisy operations), power (connectivity and controllability of every qubit) as well as the length of reliable computation (decoherence).
To work with these constraints and still achieve a quantum advantage over pure-classical computation, the quantum application community is in favour of a hybrid approach, where some parts of the computation is carried out on classical logic.

The application is modelled in the Classical Host CPU, and translated to a quantum representation using a quantum programming language (like OpenQL).
The entire application software would generally consist of one or more quantum kernels (which are suitable for acceleration) and classical pre/post processing that are required to produce the final result of the problem.
The quantum kernels are loaded to the Hybrid Quantum Accelerator using a hybrid quantum representation (like cQASM 2.0).

We consider two different types of quantum computation models for optimisation: the gate-based and the annealing-based methods.
Both models can solve an optimisation task encoded as a Quadratic Unconstrained Binary Optimisation (QUBO) model, as discussed later.

The Hybrid Quantum Accelerator typically has two processing elements as shown in Figure~\ref{f_hqa}.
The part which can benefit computationally from quantum effects like superposition, entanglement and tunnelling are offloaded to the Quantum Logic.
Since near-term quantum processors cannot run a long computation, the entire process is generally split into small chunks of quantum circuits/anneals that can be carried out in burst, measured, and restarted based on the obtained results.
The Classical Logic keeps track of this progress and suggests the quantum logic the parameters for the next trial run.

The optimisation problem is modelled as a QUBO expressed by: minimise $y = x^tQx$, where $x$ is a vector of binary decision variables ($x_i \in \{0,1\}$) and $Q$ is a (symmetric or upper triangular) square matrix of constants.
Quantum annealers use the Ising model of spin variables (with the binary variables taking the values of $\{-1,+1\}$) as the computational model.
This is isomorphic to the QUBO model and can thus be easy translated to an implementation on the annealer for estimating the minimum energy state of the spins.
QUBO models can also be solved on gate-based quantum systems using the Quantum Approximate Optimisation Algorithm (QAOA).
QAOA is a variational algorithm where the classical optimiser specifies a low-depth quantum circuit to find the lowest energy configuration of a problem Hamiltonian.
We believe that the choice of the quantum accelerator is dependent on the specific energy landscape of the application, as well as the characteristics of the quantum systems (e.g. annealers can process larger problem sizes, whereas gate models allow longer coherence times).

A specific use-case we consider here is the optimisation problem called Travelling Salesman Problem (TSP).
TSP falls under the NP-hard class (thus outside BQP), so the time to find the exact solution scales exponentially also on a quantum computer with respect to the problem size.
Often a good sub-optimal solution is admissible, thus heuristic algorithms of much lesser complexity can be employed.
Our choice of TSP is motivated by its usefulness in many industrial applications in the domains of planning, scheduling, logistics, packing, DNA sequencing, network protocols, telescope control, VLSI testing, and many more.

\begin{figure}[htb]
\centering
\includegraphics[width=0.8\textwidth]{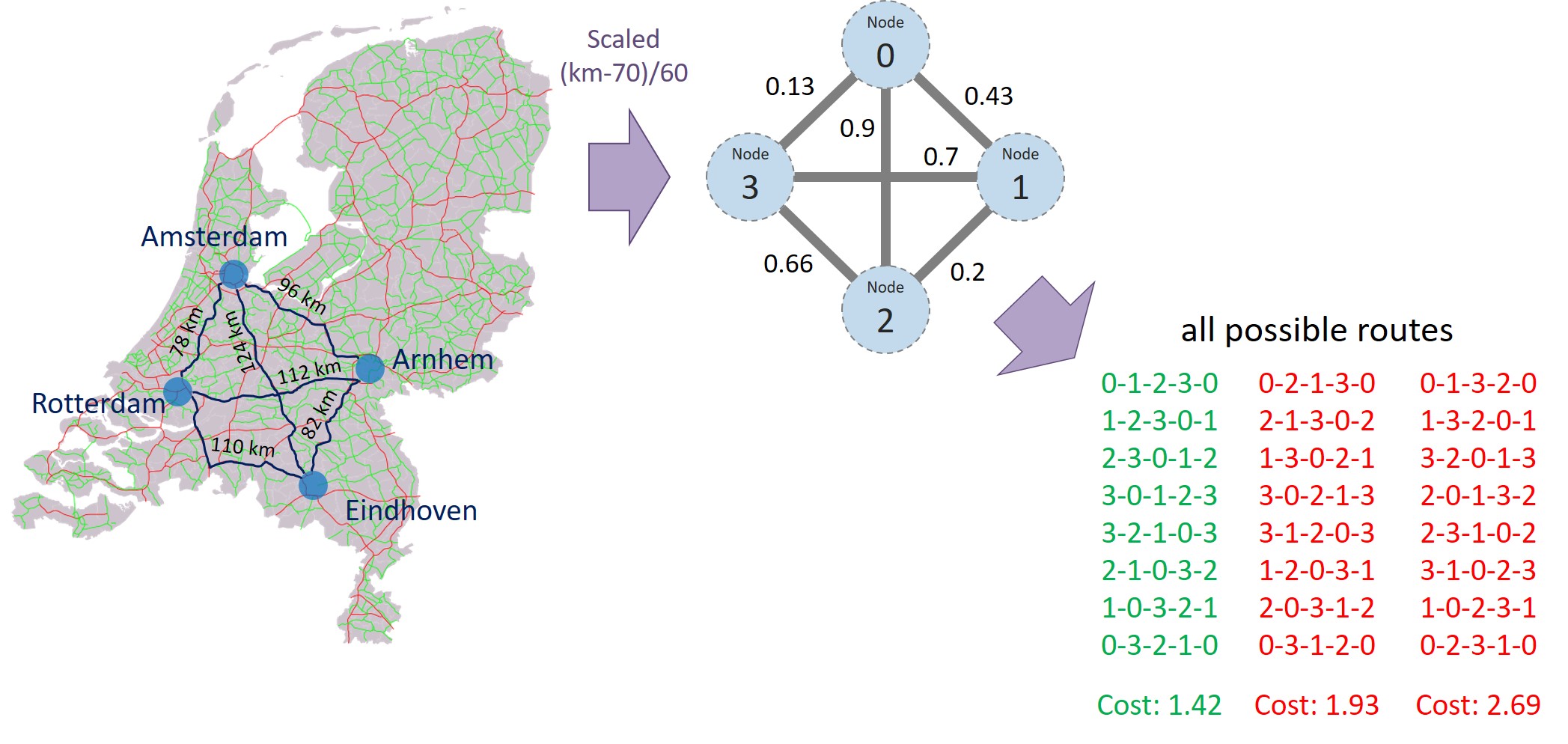}
\caption{Route-planning reduction to TSP graph}
\label{fig:qa_tsp01_new}
\end{figure}

Given a complete graph $G = (V, E)$ with weights $w_{ij}$ on the edge $i\leftrightarrow j$, the TSP aims to find a (directed/undirected) Hamiltonian cycle of minimum weight, i.e., a cycle that visits all nodes (cities) of the graph and such that the sum of the edge weights (travel cost) is minimum.
Intuitively, given the ordered pair-wise distance between cities, the TSP involves finding the shortest route that visits every city once.
The order in which these cities are visited is not constrained.
In our example, shown in Figure~\ref{fig:qa_tsp01_new}, we search the shortest route between four cities in the Netherlands.
The TSP graph is made from the scaled Euclidean distance 
We enumerate all possible solutions and find an optimal solution for this TSP with a cost of $1.42$ (as shown in green).

Since the total number of visits (time IDs) equals the total number of nodes (city IDs); the total possible combinations of '(c, t)' is square of the the number of cities.
The QUBO interactions (for the $Q$ matrix off-diagonal entries) denote pairs of 2 nodes that can/cannot coexist and the associated reward/penalty.
The interactions are categorised as follows: 
(i) every node must be assigned, 
(ii) same node assigned to two different time slot is penalised, 
(iii) same time slot assigned to two different nodes is penalised, and 
(iv) The additional cost of including an edge in the route to two consecutive time slots is the weight of the edge in the TSP tour.
We need 16 qubits to encode the example TSP into a QUBO.


When mapping the QUBO to a realistic hardware (like D-Wave 2000Q, or IBM 20 qubit System One) the connectivity of the qubits in the physical topology is important.
The embedding and mapping process considerably increases the number of required qubits and also the quality of the solution.
In the Travelling Sales Person example that was given above, the highest number of cities that can be solved on a D-Wave 2000Q machine is 9.  
The amount of qubits needed to solve the problem grows as $N^2$ and finding embedding for the case with 10 cities will fail in most (if not all) cases.
On Fujitsu's Digital Annealer, where it is fully connected (no embedding), we should be able to solve 90 cities.
Error-correction and routing for gate-based models adds further overhead in number of required qubits and operations.
In classical computation however, the current record for exact solutions to the problem, using branch and bound algorithms is 85900 cities.
Heuristics like Monte Carlo methods are used for larger inputs.
\section{Hardware and Software Long-Term Vision}


There are different ways of building a computer and the way it is currently done is to combine multiple heterogeneous multi-core processors
There are several models of quantum computation.
The theoretical models, like the quantum circuit model, adiabatic quantum computing, measurement-based (cluster state) quantum computation and topological quantum computing are equivalent to each other within polynomial time reduction.  
One of the most popular and by far the most extensively developed is the circuit model for gate-based quantum computation. 
This is the conceptual generalisation of Boolean logic gates (e.g. AND, OR, NOT, NAND, etc.) used for classical computation.
The gate set for the quantum counterpart allows a richer diversity of states on the complex vector space (Hilbert space) formed by qubit registers.
The quantum gates, by their unitary property, preserves the 2-norm of the amplitude of the states thereby undergoing a deterministic transformation of the probability distribution over bit strings.
The power of quantum computation stems from this exponential state space evolving in superposition while interacting by interference of the amplitudes.

Most of the quantum computer which are made today are based on superconducting qubits but in the past there have been attempts on ion traps and semiconducting qubits are becoming very popular.  
We are just starting to reach the 50 qubit mark in processors but are way below the required coherance.
The big system is shown in Figure~\ref{fig:layers}, where we include both the quantum annealer and the quantum gate accelerator. 
The same holds for the micro-architecture, for which, the components need to be developed.

\textbf{Full connectivity:}  An important limitation that is not yet solved in any scalable way is the connectivity between the qubits, as for two-qubit gates the qubits need to be close to each other.
It means that there is direct connectivity only in the neighbourhood of each qubits.
This has important implications on the initial mapping of the qubits on the topology and especially the routing of the qubits to a location close to the other.
Evidently, the kind of logical qubit one uses is very important. 
That is also an open issue currently brought to light by Preskill's paper~\cite{preskill2018} stating that surface codes are too expensive.
It suggests to move to small codes where much less qubits are needed to create a logical qubit.  
That is also why we have introduced the notion of a perfect qubit such that some of the complexities and problems can be abstracted away for the application developer.

Figure~\ref{fig:layers} shows our long-term schema of what a quantum computer can look like in the two directions that are currently being explored, quantum gates-based and the quantum annealing approach. 
To give an overview of what is available on the market is very difficult as there are no commercially available computer systems that can be used in any reasonable way. 
The market can be split in two parts: companies that are building a quantum-gate based computer and ones that are focusing much more on optimisation problems that can be solved with quantum annealing.

\subsection{Quantum gate-based computers}

Gate-based quantum algorithms are designed such that the solution states interfere constructively while the non-solutions interfere destructively, biasing the final probability distribution in favour of reading out the solution(s).  However, the error rates are still around $10^{-2/-3}$ and need to be substantially improved.

\begin{itemize}[nolistsep,noitemsep]
    \item IBM: they offer a quantum processor up to 53 qubits. The qubits have all the normal error behaviour but they can be programmed. The specific thing is that IBM has not yet looked at any micro-architectural control of the physical level.
    \item Intel: is looking at both semi- and superconducting qubits but are in essence more interested in the semi-conducting qubit processor. The essence is fixing a lot on the qubit production, partly supported by a solid micro-architecture.
    \item Microsoft: has some preference for the majorana-based approach but they still have to make the first qubit based on that quasi-particle. They are very active in the software development.
    \item Alibaba: is a strong player in this field and they have a quantum lab that focuses on a range of activities going from the development of a quantum processor, quantum-classical algorithms up to simulation of quantum physics.
    \item Google: also one of the leaders in superconducting qubits (John Martinis's team)
    \item Rigetti: is a start-up 
    focusing on the superconducting quantum processor. They advance well but there is not yet any applicable processor in the market even though there is a processor that can be used for some testing purposes.
    
    \item Xanadu: the team focuses on continuous variable quantum computing based on photonics of squeezed light.
\end{itemize}

\subsection{Quantum annealing-based computers}

Quantum annealing has a slightly different software stack than gate-model quantum computers and must be interpreted as a  more limited edition of a quantum accelerator based on quantum gates algorithms.
Instead of a quantum circuit, the level of abstraction is the classical Ising model, i.e. the problem we are interested in solving must be in this form.  
Just like superconducting gate-model quantum computers, superconducting quantum annealers also suffer from limited connectivity.
It means that we have to find a graph minor embedding, combining several physical qubits into a logical qubit.  
inding an embedding is NP-hard in itself, so probabilistic heuristics are normally used. 
\footnote{Reference to the QAnnealing workflow: Open source software in quantum computing - arxiv.org/abs/1812.09167}
We make a distinction between companies that offer a quantum computer such as D-Wave or a quantum-inspired computer such as Fujitsu. QNNcloud is a third offer based on neural-network and optical based quantum mechanisms.

\begin{itemize}[nolistsep,noitemsep]
    \item D-Wave: The technology is up to 2000 superconducting qubits (in 2018), compared to the less than 100  qubits on gate-model quantum computers. D-Wave Systems has been building superconducting quantum annealers for over a decade. D-Wave Systems company offers an open source suite called Ocean which can be used to make small examples of applications which can be executed on a D-Wave computer.
    
    \item Fujitsu: has invested in the development of a digital annealer. They offer a quantum-inspired computer and not a quantum computer. It is meant for the same kind of optimisation problems that D-Wave can handle (QUBO problems). They currently offer 8192 nodes with full-connectivity and a programming interface but it is not clear and not known how the quantum-inspired accelerator works.
    
    \item Hitachi: Similar to Fujitsu, Hitachi is also specialising in making a quantum accelerator based on quantum annealing using semiconducting qubits.  More information can be found on the URL site mentioned here.
    \footnote{$ https://www.hitachi.com/rd/portal/contents/story/cmos_annealing2/index.html$}
    
    \item QNNcloud: is a company that offers a neural network based optical quantum computer where the neurons can be put in superposition and quantum-measurement circuits. A quantum optics implementation by QNNcloud uses a coherent Ising model, having different restrictions from superconducting architectures.
    
    \item 1QBit: develops general purpose algorithms for quantum computing hardware, primarily focused on computational finance, materials science, quantum chemistry, and the life sciences. While there is a plethora of quantum computing languages, frameworks, and libraries for the gate-model, quantum annealing is less well-established. Their 1Qloud platform is focused on mapping optimisation problems into QUBO format necessary to compute with quantum annealing processors and similar devices from Fujitsu, D-Wave, Hitachi and NTT (QNNcloud), while their QEMIST platform is focused on advanced materials and quantum chemistry research with universal quantum computing processors.
\end{itemize}

\subsection{Quantum programming languages}

A last component of the offering is related to the programming language that can be used.

\begin{itemize}[nolistsep,noitemsep]
    \item Qiskit: is IBM's open-source quantum computing software development framework for leveraging available quantum processors. It consists of Terra (core compiler and libraries for quantum programming), Aer (noise modelling and noise-free simulators), Ignis (characterisation of errors and QEC) and Aqua (applications).

    \item Forest: is Rigetti's SDK includes a simulator and cloud connection utilities.
    
    \item CirQ: is Google's open-source quantum framework for experimenting with noisy intermediate scale quantum (NISQ) algorithms on near-term quantum processors. Also, the OpenFermion platform helps in translating problems in chemistry and materials science into quantum circuits.

    \item Strawberry Fields: is Xanadu's full-stack Python library for designing, simulating, and optimising quantum optical circuits. Xanadu also offers the Blackbird quantum programming language and the PennyLane quantum machine learning platform.
    
    \item XACC: A vendor-independent solution is XACC, an extensible compilation framework for hybrid quantum-classical computing architectures.
    
    \item OpenJij: is a open framework for the Ising model and QUBO. 
    
    \item QMASM: is a quantum macro assembler for D-Wave systems from Los Alamos National Laboratories. It fills a gap in the software ecosystem for D-Wave's adiabatic quantum computers by shielding the programmer from having to know system-specific hardware details while still enabling programs to be expressed at a fairly low level of abstraction. 
    
   \item OpenQL: that is the language which was discussed in this paper earlier.
\end{itemize}

\section{Towards In-Memory Computing}

In-memory computing is becoming increasingly important as a new computer architecture. 
Rather than moving the huge amounts of data around to the logic, it is much more meaningful to move the logic around and keep the data as local as possible without moving it around, using, for instance, innovative technology such as memristors. 
Memristors were theoretically defined already several decades ago by Leon Chua, but recently the semiconductor manufacturers are seriously investigating their production.
The key idea of a memristor is that it can be used to store data but also to make calculations. 
This is why memristors are an ideal candidate for making an in-memory architecture.
The concept of in-memory computing is described in a paper where the concept is illustrated using memristor based devices~\cite{hamdioui2015}. 
The main advantage of memristors is that they can be used both to store information and to work on it. 
So an intelligent merging of logic with data storage is the key of an in-memory architecture.  
It is a completely new way of designing algorithms and computing systems and it is far from evident what the design rules are that are needed to fully exploit the in-memory computing potential. 

The link with quantum computing is very straight: the quantum logic is directly applied on the qubits and the qubits do not need to be transported to any Quantum Arithmetic and Logical Unit (ALU) before being  processed.   
In quantum computing, the routing of qubit states is therefore also a very important problem. 
The qubits need to be put on the quantum chip in a way that the movement of qubit states is as minimal as possible. 
Also what routing protocols will be used for any quantum chip is a big open area of research in quantum computer engineering. 
Currently, in any of the semiconducting or superconducting quantum implementations the interaction between qubits has a nearest-neighbour constraint. 
That induces the need for deciding where to map and how to route the qubits used in the algorithm on the quantum chip. 
This qubit routing is an important and illustrative example of what in-memory quantum computing actually means.
When adopting an in-memory computing architecture, a crucial challenge is to decide on the placement of the data that needs to be processed and to have a programming language and compiler, such that the appropriate logic can placed close to the data. 
Any kind of algorithm will have data that the algorithm is changing to get a result and it is quite unlikely that there is no dependency between any of those data items. 
What that implies is that intermediate results will have to move around in the architecture such that it reaches the place where that result is used in the next computational step. 
Even though in-memory puts all the data in some kind of memory, those data items still have to move around such that a final result can be computed by the classical Host-CPU.  
From a quantum physics point of view, the main challenges are the coherence of the qubits, the fidelity of the operations and the overall error rate of the quantum computation, involving both the qubits as well as their operation and the involved error-corrections.  
This is already being sufficiently studied by the quantum community but there are also clearly other challenges that need to be researched as soon as possible.

One of the main problems is the error-proneness of the qubit behaviour which consumes up to 90\% of the (quantum) computer time. 
As explained, the routing and moving around of qubit states is a very important challenge.  
So any progress the physics community is making in that respect is extremely important as it will reduce substantially the pressure on the micro-architecture and the overall system design.  
In \cite{brennen2003quantum}, the authors present a quantum computer architecture that addresses the important problem of qubit state routing for nearest-neighbour two-qubit gate execution. 
They use an idea from the von-Neumann architecture of classical machines such as a quantum bus which is a refreshable entanglement resource to connect distant memory nodes. 
The overall approach is at the level of entanglement purification and qubit pairs with different fidelities.
Given that a quantum computation on qubits complies to the same overall in-memory computing logic, that particular architecture is definitely interesting for any quantum device.  
The challenges involved with in-memory computing are therefore the same as for quantum computing. The underlying technology are not memristors or other technology but any of the quantum technologies and require also a full-stack integration of the different layers.  
In that sense, the quantum computing research should be based very much on the basic principles of in-memory computing. 
\section{Future Prospects}

\begin{figure}[htb] 
    \centering
    \captionsetup{justification=centering}
    \subfigure[Development time frame]
    { 
        \centering
        \includegraphics[width=.56\textwidth]{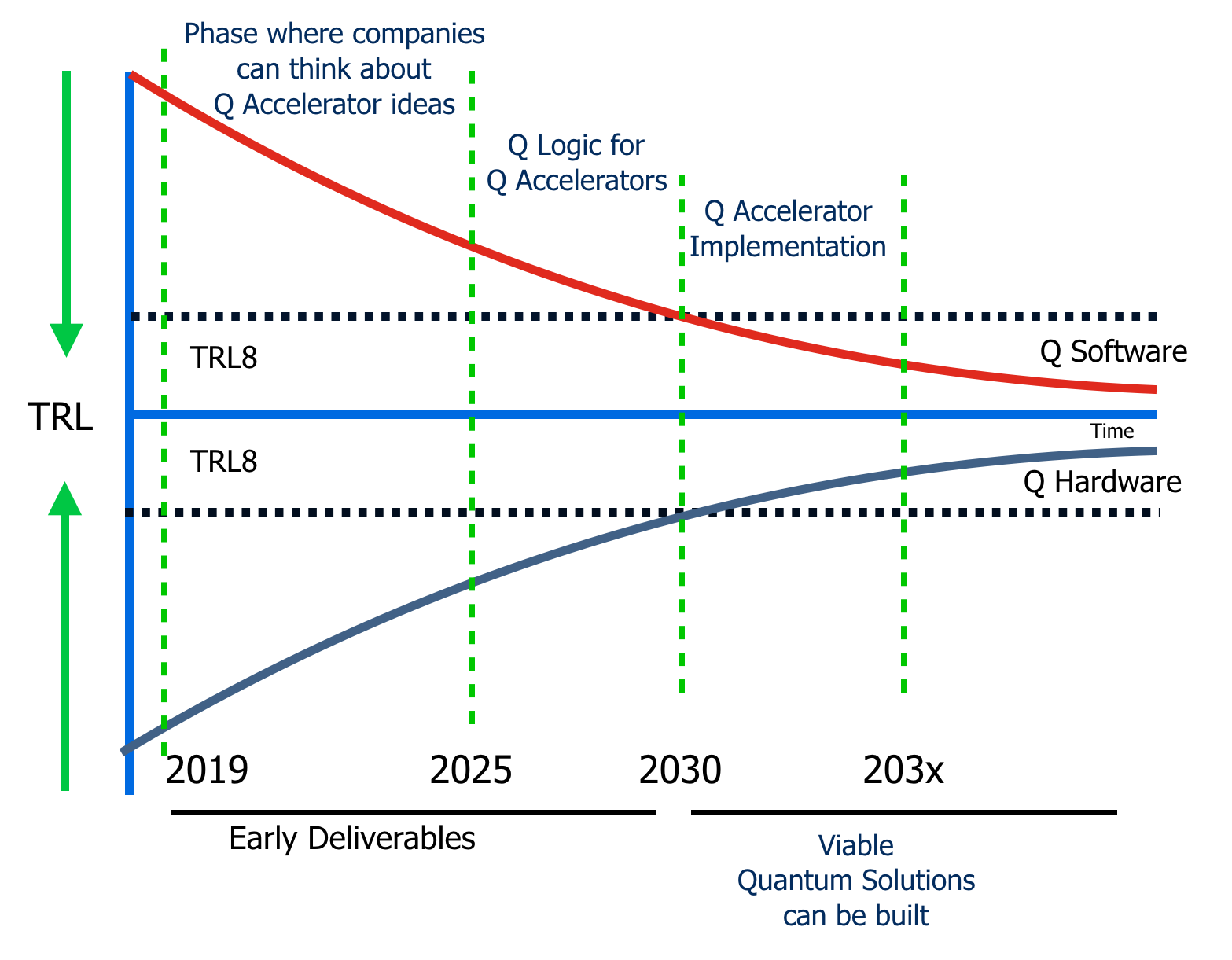}
        \label{fig:time-frame} 
    } 
    \subfigure[Structural division between perfect and realistic qubits]
    { 
        \centering
        \includegraphics[width=.36\textwidth]{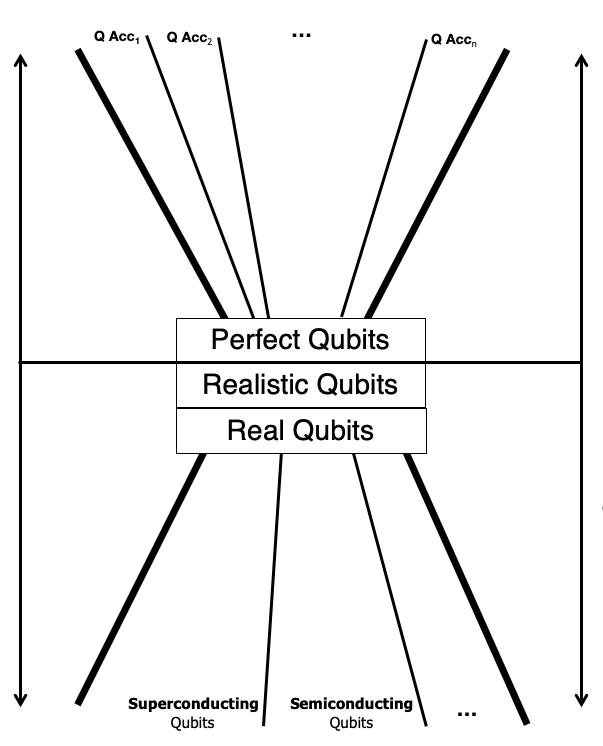}
        \label{fig:structural-division}
    }
    \caption{Quantum computer development future projections}
    \label{fig:future-projections}
\end{figure}

It is very important that companies and other organisations start investing as soon as possible in Quantum Technology.
Figure~\ref{fig:future-projections} shows a projection of when different parts of software and hardware development will be required, to create an efficient quantum computer.
The distinction is made between the use of quantum accelerators and that of manufacturing a quantum chip.  
In general, any commercial or other organisation is interested in new technology if the Technology Readiness Level~(TRL) is high enough.  
If we adopt the same levels as for classical technology, the TRL needs to have reached level 8 and that is sketched in the red and black line that are shown in Figure~\ref{fig:time-frame}.  
There are 4 vertical, green-dotted lines to illustrate 3 moments leading to the last phase where we assume there is enough software or hardware maturity that can be used for any accelerator one wants to build.  
Phase I focuses on the reflection by the organisation on the concrete need that exists and for which a quantum accelerator logic can be developed.   
Phase II resembles the team members brainstorming on the logic for the quantum accelerator. 
They will express that logic in OpenQL and develop some prototype micro-architecture and executed the logic on the QX-simulator.  
Phase III then focuses exclusively on the actual implementation and execution of the Quantum Accelerator logic, whether on an experimental quantum chip or on the QX simulator. 
This is the moment when the top and low curves can be combined in a real quantum prototype of the accelerator.  
Figure~\ref{fig:structural-division} represents the way that the two lines of research are currently separated and which will be joined in maybe over the next decade.
The division was used in this paper where we made the distinction between the use of perfect and realistic qubits and how that determines the different layers in the full-stack.
\section{Conclusion}

Over the last couple of decades, quantum computing has been a one-dimensional research effort focusing on understanding how to make coherent qubits and how to implement the different universal quantum gate sets on any of the multiple quantum approaches.
As far as computer architectural choices were made, the community has been focused very much on the von-Neumann computer architecture and defined qubits in terms of memory and processing qubits.
However, computer engineering as a field has understood by now that this approach never scales to the size needed for handling, for  instance, the Big Data volumes that world wide are being generated and collected.
Two approaches seem to be very promising: the first comes from the accelerator community and involves the full stack integration of the different layers that are needed to build the quantum accelerator.
The use of perfect qubits in that context makes sense as the end-users of any quantum accelerator can focus their reasoning on the quantum logic of the application and verify it through some implementation of the micro-architecture and the execution of the quantum instructions on the quantum simulator.
The second option is to use the full-stack for the control of, for instance, superconducting and semiconducting qubits with a micro-code layer where we translate any kind of common QASM into an operational set of micro-instructions, for a meaningful adoption of existing computer technology.
It is very difficult to predict the performance improvement of a quantum computational device but that it will be much higher than any existing computational technology is clear.
It also depends on the quantum application that is being looked at and the way the qubits are manufactured.
Research is still needed for at least a decade before the full-integration effects become visible and verifiable.


\bibliographystyle{unsrt}       
\bibliography{refs}

\begin{thebibliography}{10}
\expandafter\ifx\csname url\endcsname\relax
  \def\url#1{\texttt{#1}}\fi
\expandafter\ifx\csname urlprefix\endcsname\relax\def\urlprefix{URL }\fi
\providecommand{\bibinfo}[2]{#2}
\providecommand{\eprint}[2][]{\url{#2}}

\bibitem{vassiliadis2004}
\bibinfo{author}{Vassiliadis, S.} \emph{et~al.}
\newblock \bibinfo{title}{The molen polymorphic processor}.
\newblock \emph{\bibinfo{journal}{IEEE Transactions on Computers}}
  \textbf{\bibinfo{volume}{53}}, \bibinfo{pages}{1363--1375}
  (\bibinfo{year}{2004}).

\bibitem{preskill2018quantum}
\bibinfo{author}{Preskill, J.}
\newblock \bibinfo{title}{Quantum computing in the {NISQ} era and beyond}.
\newblock \emph{\bibinfo{journal}{arXiv:1801.00862}}  (\bibinfo{year}{2018}).

\bibitem{feynman1982simulating}
\bibinfo{author}{Feynman, R.~P.}
\newblock \bibinfo{title}{Simulating physics with computers}.
\newblock \emph{\bibinfo{journal}{International Journal of Theoretical
  Physics}} \textbf{\bibinfo{volume}{{21}}}, \bibinfo{pages}{{467--488}}
  (\bibinfo{year}{{1982}}).

\bibitem{van2013blueprint}
\bibinfo{author}{Van~Meter, R.} \& \bibinfo{author}{Horsman, C.}
\newblock \bibinfo{title}{A blueprint for building a quantum computer}.
\newblock \emph{\bibinfo{journal}{Communications of the ACM}}
  \textbf{\bibinfo{volume}{56}}, \bibinfo{pages}{84--93}
  (\bibinfo{year}{2013}).

\bibitem{divincenzo2000physical}
\bibinfo{author}{DiVincenzo, D.~P.}
\newblock \bibinfo{title}{The physical implementation of quantum computation}.
\newblock \emph{\bibinfo{journal}{arXiv preprint quant-ph/0002077}}
  (\bibinfo{year}{2000}).

\bibitem{preskill2018}
\bibinfo{author}{Preskill, J.}
\newblock \bibinfo{title}{Quantum computing in the nisq era and beyond}.
\newblock \emph{\bibinfo{journal}{Arxiv}}
  \textbf{\bibinfo{volume}{1801.00862}}, \bibinfo{pages}{20}
  (\bibinfo{year}{2018}).

\bibitem{shor1994algorithms}
\bibinfo{author}{Shor, P.~W.}
\newblock \bibinfo{title}{Algorithms for quantum computation: discrete
  logarithms and factoring}.
\newblock In \emph{\bibinfo{booktitle}{Foundations of Computer Science, 1994
  Proceedings., 35th Annual Symposium on}}, \bibinfo{pages}{124--134}
  (\bibinfo{year}{1994}).

\bibitem{zalka1999grover}
\bibinfo{author}{Zalka, C.}
\newblock \bibinfo{title}{Grover’s quantum searching algorithm is optimal}.
\newblock \emph{\bibinfo{journal}{Physical Review A}}
  \textbf{\bibinfo{volume}{60}}, \bibinfo{pages}{2746} (\bibinfo{year}{1999}).

\bibitem{houtgast2018hardware}
\bibinfo{author}{Houtgast, E.~J.}, \bibinfo{author}{Sima, V.-M.},
  \bibinfo{author}{Bertels, K.} \& \bibinfo{author}{Al-Ars, Z.}
\newblock \bibinfo{title}{Hardware acceleration of bwa-mem genomic short read
  mapping for longer read lengths}.
\newblock \emph{\bibinfo{journal}{Computational biology and chemistry}}
  \textbf{\bibinfo{volume}{75}}, \bibinfo{pages}{54--64}
  (\bibinfo{year}{2018}).

\bibitem{svore2018q}
\bibinfo{author}{Svore, K.} \emph{et~al.}
\newblock \bibinfo{title}{Q\#: Enabling scalable quantum computing and
  development with a high-level dsl}.
\newblock In \emph{\bibinfo{booktitle}{Proceedings of the Real World Domain
  Specific Languages Workshop 2018}}, \bibinfo{pages}{7}
  (\bibinfo{organization}{ACM}, \bibinfo{year}{2018}).

\bibitem{abhari2012scaffold}
\bibinfo{author}{Abhari, A.~J.} \emph{et~al.}
\newblock \bibinfo{title}{Scaffold: Quantum programming language}.
\newblock \bibinfo{type}{Tech. Rep.}, \bibinfo{institution}{Princeton
  University} (\bibinfo{year}{2012}).

\bibitem{green2013introduction}
\bibinfo{author}{Green, A.~S.}, \bibinfo{author}{Lumsdaine, P.~L.},
  \bibinfo{author}{Ross, N.~J.}, \bibinfo{author}{Selinger, P.} \&
  \bibinfo{author}{Valiron, B.}
\newblock \bibinfo{title}{An introduction to quantum programming in quipper}.
\newblock In \emph{\bibinfo{booktitle}{International Conference on Reversible
  Computation}}, \bibinfo{pages}{110--124} (\bibinfo{organization}{Springer},
  \bibinfo{year}{2013}).

\bibitem{khammassi18}
\bibinfo{author}{Khammassi, N.} \emph{et~al.}
\newblock \bibinfo{title}{Openql 1.0: A quantum programming language for
  quantum accelerators,}.
\newblock \emph{\bibinfo{journal}{QCA Technical Report}} \bibinfo{pages}{8}
  (\bibinfo{year}{2018}).

\bibitem{DiCarlo2015}
\bibinfo{author}{Riste, D.}, \bibinfo{author}{Poletto, S.},
  \bibinfo{author}{Huang, M.~Z.} \emph{et~al.}
\newblock \bibinfo{title}{Detecting bit-flip errors in a logical qubit using
  stabilizer measurements}.
\newblock \emph{\bibinfo{journal}{Nat Commun}} \textbf{\bibinfo{volume}{6}}
  (\bibinfo{year}{2015}).
\newblock \urlprefix\url{http://dx.doi.org/10.1038/ncomms7983}.

\bibitem{corcoles2015demonstration}
\bibinfo{author}{C{\'o}rcoles, A.} \emph{et~al.}
\newblock \bibinfo{title}{Demonstration of a quantum error detection code using
  a square lattice of four superconducting qubits}.
\newblock \emph{\bibinfo{journal}{Nature communications}}
  \textbf{\bibinfo{volume}{6}} (\bibinfo{year}{2015}).

\bibitem{Martinis2015}
\bibinfo{author}{Kelly, J.} \emph{et~al.}
\newblock \bibinfo{title}{State preservation by repetitive error detection in a
  superconducting quantum circuit}.
\newblock \emph{\bibinfo{journal}{Nature}} \textbf{\bibinfo{volume}{519}},
  \bibinfo{pages}{66--69} (\bibinfo{year}{2015}).

\bibitem{lidar}
\bibinfo{author}{Lidar~D., B.~T.}
\newblock \emph{\bibinfo{title}{Quantum Error Correction}}
  (\bibinfo{publisher}{Cambridge university press}, \bibinfo{year}{2013}).

\bibitem{shor1995scheme}
\bibinfo{author}{Shor, P.~W.}
\newblock \bibinfo{title}{Scheme for reducing decoherence in quantum computer
  memory}.
\newblock \emph{\bibinfo{journal}{Physical review A}}
  \textbf{\bibinfo{volume}{52}}, \bibinfo{pages}{R2493} (\bibinfo{year}{1995}).

\bibitem{steane1996multiple}
\bibinfo{author}{Steane, A.}
\newblock \bibinfo{title}{Multiple-particle interference and quantum error
  correction}.
\newblock In \emph{\bibinfo{booktitle}{Proceedings of the Royal Society of
  London A: Math., Phys. and Eng. Sciences}} (\bibinfo{year}{1996}).

\bibitem{calderbank1996good}
\bibinfo{author}{Calderbank, A.~R.} \& \bibinfo{author}{Shor, P.~W.}
\newblock \bibinfo{title}{Good quantum error-correcting codes exist}.
\newblock \emph{\bibinfo{journal}{Phys. Rev. A}} \textbf{\bibinfo{volume}{54}},
  \bibinfo{pages}{1098} (\bibinfo{year}{1996}).

\bibitem{gottesman1996class}
\bibinfo{author}{Gottesman, D.}
\newblock \bibinfo{title}{Class of quantum error-correcting codes saturating
  the quantum hamming bound}.
\newblock \emph{\bibinfo{journal}{Phys. Rev. A}} \textbf{\bibinfo{volume}{54}},
  \bibinfo{pages}{1862} (\bibinfo{year}{1996}).

\bibitem{bombin2006topological}
\bibinfo{author}{Bombin, H.} \& \bibinfo{author}{Martin-Delgado, M.~A.}
\newblock \bibinfo{title}{Topological quantum distillation}.
\newblock \emph{\bibinfo{journal}{Phys. Rev. Lett.}}
  \textbf{\bibinfo{volume}{97}}, \bibinfo{pages}{180501}
  (\bibinfo{year}{2006}).

\bibitem{fowler2012surface}
\bibinfo{author}{Fowler, A.~G.}, \bibinfo{author}{Mariantoni, M.},
  \bibinfo{author}{Martinis, J.~M.} \& \bibinfo{author}{Cleland, A.~N.}
\newblock \bibinfo{title}{Surface codes: Towards practical large-scale quantum
  computation}.
\newblock \emph{\bibinfo{journal}{Physical Review A}}
  \textbf{\bibinfo{volume}{86}}, \bibinfo{pages}{032324}
  (\bibinfo{year}{2012}).

\bibitem{lin2015paqcs}
\bibinfo{author}{Lin, C.-C.} \emph{et~al.}
\newblock \bibinfo{title}{Paqcs: Physical design-aware fault-tolerant quantum
  circuit synthesis}.
\newblock \emph{\bibinfo{journal}{IEEE Transactions on VLSI Systems}}
  \textbf{\bibinfo{volume}{23}}, \bibinfo{pages}{1221--1234}
  (\bibinfo{year}{2015}).

\bibitem{dousti12min}
\bibinfo{author}{Dousti, M.~J.} \& \bibinfo{author}{Pedram, M.}
\newblock \bibinfo{title}{Minimizing the latency of quantum circuits during
  mapping to the ion-trap circuit fabric}.
\newblock In \emph{\bibinfo{booktitle}{DATE}} (\bibinfo{year}{2012}).

\bibitem{horsman2012surface}
\bibinfo{author}{Horsman, C.}, \bibinfo{author}{Fowler, A.~G.},
  \bibinfo{author}{Devitt, S.} \& \bibinfo{author}{Van~Meter, R.}
\newblock \bibinfo{title}{Surface code quantum computing by lattice surgery}.
\newblock \emph{\bibinfo{journal}{New Journal of Physics}}
  \textbf{\bibinfo{volume}{14}}, \bibinfo{pages}{123011}
  (\bibinfo{year}{2012}).

\bibitem{fu2018eqasm}
\bibinfo{author}{Fu, X.} \emph{et~al.}
\newblock \bibinfo{title}{{eQASM}: An executable quantum instruction set
  architecture}.
\newblock \emph{\bibinfo{journal}{arXiv:1808.02449}}  (\bibinfo{year}{2018}).

\bibitem{grover1997quantum}
\bibinfo{author}{Grover, L.~K.}
\newblock \bibinfo{title}{Quantum mechanics helps in searching for a needle in
  a haystack}.
\newblock \emph{\bibinfo{journal}{Physical review letters}}
  \textbf{\bibinfo{volume}{79}}, \bibinfo{pages}{325} (\bibinfo{year}{1997}).

\bibitem{ventura}
\bibinfo{author}{Wang, D. V.~P.} \emph{et~al.}
\newblock \bibinfo{title}{Artificial associative memory using quantum
  processes}.
\newblock \emph{\bibinfo{journal}{Proceed. Joint Conf. on Information Sci.}}
  \textbf{\bibinfo{volume}{2}}, \bibinfo{pages}{218--221}
  (\bibinfo{year}{1998}).

\bibitem{AritraMSc}
\bibinfo{author}{Sarkar, A.}
\newblock \emph{\bibinfo{title}{MSc thesis : Quantum Algorithms for
  pattern-matching in genomic sequences}} (\bibinfo{year}{2018}).

\bibitem{sarkar2019algorithm}
\bibinfo{author}{Sarkar, A.}, \bibinfo{author}{Al-Ars, Z.},
  \bibinfo{author}{Almudever, C.~G.} \& \bibinfo{author}{Bertels, K.}
\newblock \bibinfo{title}{An algorithm for dna read alignment on quantum
  accelerators}.
\newblock \emph{\bibinfo{journal}{arXiv preprint arXiv:1909.05563}}
  (\bibinfo{year}{2019}).

\bibitem{svore2006layered}
\bibinfo{author}{Svore, K.~M.}, \bibinfo{author}{Aho, A.~V.},
  \bibinfo{author}{Cross, A.~W.}, \bibinfo{author}{Chuang, I.} \&
  \bibinfo{author}{Markov, I.~L.}
\newblock \bibinfo{title}{A layered software architecture for quantum computing
  design tools}.
\newblock \emph{\bibinfo{journal}{Computer}} \bibinfo{pages}{74--83}
  (\bibinfo{year}{2006}).

\bibitem{haner2016software}
\bibinfo{author}{H{\"a}ner, T.}, \bibinfo{author}{Steiger, D.~S.},
  \bibinfo{author}{Svore, K.} \& \bibinfo{author}{Troyer, M.}
\newblock \bibinfo{title}{A software methodology for compiling quantum
  programs}.
\newblock \emph{\bibinfo{journal}{arXiv preprint arXiv:1604.01401}}
  (\bibinfo{year}{2016}).

\bibitem{hamdioui2015}
\bibinfo{author}{Hamdioui, S.} \emph{et~al.}
\newblock \bibinfo{title}{Memristor based computation-in-memory architecture
  for data-intensive applications}.
\newblock In \emph{\bibinfo{booktitle}{2015 Design, Automation Test in Europe
  Conference Exhibition (DATE)}}, \bibinfo{pages}{1718--1725}
  (\bibinfo{year}{2015}).

\bibitem{brennen2003quantum}
\bibinfo{author}{Brennen, G.~K.}, \bibinfo{author}{Song, D.} \&
  \bibinfo{author}{Williams, C.~J.}
\newblock \bibinfo{title}{Quantum-computer architecture using nonlocal
  interactions}.
\newblock \emph{\bibinfo{journal}{Physical Review A}}
  \textbf{\bibinfo{volume}{67}}, \bibinfo{pages}{050302}
  (\bibinfo{year}{2003}).

\end{thebibliography}

\end{document}